\documentclass[11pt,a4wide]{article}
\usepackage[margin=1.5in]{geometry}
\usepackage[utf8]{inputenc}
\usepackage{amsmath,amsthm}
\usepackage{amssymb}
\usepackage{graphicx}
\usepackage{hyperref}
\usepackage{floatrow}
\usepackage{booktabs}
\usepackage{adjustbox}
\usepackage{authblk}
\usepackage{lscape}
\usepackage{threeparttable}
\usepackage[official]{eurosym}
\usepackage{mwe}
\usepackage{acro}
\usepackage[toc,page]{appendix}
\usepackage{caption}
\usepackage{subcaption}
\usepackage{lineno}
\usepackage{siunitx}
\usepackage{comment}
\usepackage{dsfont}
\usepackage{xcolor}
\usepackage{makecell}
\usepackage{multirow}
\usepackage[numbers]{natbib}
\usepackage[noabbrev, capitalise]{cleveref}
\usepackage{tabularx}
\usepackage{colortbl}
\usepackage{tabulary}
\usepackage{etoolbox}
\usepackage{rotating}
\usepackage{bbding}
\usepackage{pgfplots}
\usepackage{enumitem}
\usepackage{tikzsymbols}
\usepackage{threeparttable}
\usepackage{slashbox}
\usepackage{makecell}
\usepackage{tikz}
\usetikzlibrary{arrows.meta, automata,
                positioning,
                quotes}
\tikzset{mynode/.style={draw, very thick, circle, minimum size=1cm},
    }
\tikzset{>=Latex} 

\acsetup{first-style=short}

\newcommand{\set}[1]{\left\{#1\right\}}
\newcommand{\abs}[1]{\left\vert#1\right\vert}
\newcommand{\Real}{\mathds R}
\newtheorem{thm}{Theorem}[]

\theoremstyle{definition}

\newtheorem{ex}[thm]{Example}
\newcommand{\BL}{BL}
\newcommand{\DP}{D\!P}

\title{A model-based analysis of the AggregateEU mechanism:
Implications of overbidding and non-commitment}

\author[1,2]{D\'{a}vid Csercsik\footnote{\textit{Corresponding author:} email; \textit{Address:} Institute of Economics, HUN-REN Centre for Economic and Regional Studies, Hungary}}
\author[3]{Borb\'{a}la Tak\'{a}csn\'{e} T\'{o}th}
\author[4]{P\'{e}ter Kotek}
\author[1,5]{L\'{a}szl\'{o} \'{A}. K\'{o}czy}
\author[6]{Anne Neumann}

\affil[1]{Institute of Economics, HUN-REN Centre for Economic and Regional Studies, Hungary}
\affil[2]{Department of Information Technology and Bionics, P\'{a}zm\'{a}ny P\'{e}ter Catholic University, Hungary}
\affil[3]{Corvinus University of Budapest, Hungary}
\affil[4]{Regional Centre for Energy Policy Research, Budapest, Hungary}
\affil[5]{Department of Finance, Budapest University of Technology and Economics, Budapest, Hungary}
\affil[6]{Department of Industrial Economics and Technology Management, \protect\\ The Norwegian University of Science and Technology, Trondheim, Norway}

\date{}

\providecommand{\keywords}[1]{\textbf{\textit{Keywords---}} #1}
\begin{document}

\maketitle
\begin{abstract}
AggregateEU is a new centralised mechanism that provides a no-commitment platform to trade natural gas in the European Union. Throughout the consultation process, AggregateEU has been mocked as `Tinder of the European gas markets' as it helps consumers and suppliers find partners, but leaves it up to the matched partners to decide whether to contract for potential trade. The non-commitment nature leads to substantial overbidding and many non-realised matches.

We propose a quantitative modelling framework to study the effect of overbidding in the AggergateEU demand aggregation or joint purchasing mechanism. We conclude that the mechanism is prone to overbidding and that overbidding has ambiguous effects on trade. Depending on the parameters, overbidding may facilitate trade, but may also result in highly inefficient outcomes when overbidding is combined with miscoordination over the delivery points.

Suggested remedies include allowing for convex bids, restrictions on overbidding, or giving up part of the non-binding character of the market.
Our results suggest that a potential future mechanism allowing the coordination of multiple delivery points could enhance the efficiency of gas markets.

\end{abstract}

\maketitle

\newpage
\section{Introduction}\label{sec:intro}
The 2021–2022 European energy crisis exposed the vulnerability of energy supply and Europe's dependence on Russian natural gas. Russian pipeline gas accounted for approximately 40\% of the EU27’s natural gas consumption, supplying about 160 bcm/year before the crisis. However, Russian gas supplies already started declining in 2021, before Russia’s invasion of Ukraine.
In 2021, the Yamal pipeline ceased deliveries via Poland, resulting in an unprecedented surge in European natural gas prices. Prices surged to \euro 100/MWh —a sharp increase from the usual \euro 25/MWh— and, for the first time in history, exceeded Asian natural gas prices. During the first half of 2022, as Russian gas deliveries continued to drop, European spot prices on the Title Transfer Facility (TTF) rose even further.
In May 2022, the EU introduced the REPowerEU strategy \citep{eurepower2022}, aiming to phase out Russian natural gas by 2027 and reduce Russian natural gas imports by 75\% by the end of 2022. The strategy outlined key measures such as:
\begin{itemize}
    \item replacing pipeline gas with LNG,
    \item accelerating the development of missing infrastructure,
    \item implementing energy-saving measures and efficiency improvements\footnote{Piebalgs and Olczak \cite{piebalgs2023eu} add curbing methane emissions as a possiblity.},
    \item substituting natural gas with renewable energy sources.
\end{itemize}
In addition, the EU/2022/1032 Gas Storage Regulation \citep{eureg2022_1032} required EU countries to fill their gas storage facilities to 80\% of their capacity in preparation for winter 2022/23 and 90\% for subsequent winters. Since storage levels were initially very low, as market players were reluctant to stockpile gas, this obligation triggered an enormous short-term surge in demand during the summer of 2022. Combined with supply shortages and reduced trading volumes, TTF gas prices skyrocketed to over \euro 300/MWh. However, once the storage target was reached by August 2022, prices dropped sharply, stabilising between \euro 25-€50/MWh.
In its ex-post evaluation of the 2022 price spike, ACER (the EU Agency for the Cooperation of Energy Regulators) \citep{ACER2023} finds that the storage requirement had intensified competition among European buyers, further driving up prices. As a result, voices demanding change in the EU's energy policy have strengthened \citep{misik2022eu}.

To address the challenges, the European Commission (EC) proposed the Aggregate EU Mechanism (Council Regulation 2022/2576)\citep{EU2022reg_aggreu} that came into force in December 2022.
The core idea of AggregateEU is to create a platform for European buyers to pool their demand for the upcoming winters, acting as an anonymous counterparty to increase coordination and strengthen their bargaining power against sellers. To trigger liquidity, a minimum volume requirement was set for EU member states to bid for at least 15\% of their yearly storage obligation via the platform. On the supply side, Russian natural gas was explicitly banned from the platform.
According to the EC's perspective, while the mechanism does not seek to replace existing trading platforms in the EU, it operates without financial clearing, making participation easier and potentially more cost-effective.
As a result, matches on the platform did not typically lead to immediate contracts, unlike those on an exchange. Instead, the platform merely informed parties of a match, leaving them to negotiate and finalise contracts independently.
According to ACER, the AggregateEU joint purchasing platform will be expanded in the future to also include renewable gases, such as biomethane \citep{ACER2025}.

\subsection{Related literature}

Various authors have already researched group buying. In one of the first pioneering works \cite{anand2003group}, Anand and Aron approach the question of group buying from the perspective of firms selling a product under demand uncertainty and characterize the optimal group buying schedule.
While Dana argues that group buying may be beneficial for buyers with heterogeneous preferences even in the case of small groups \cite{dana2012buyer}, the paper \cite{chen2011group} points out that the concept may be detrimental for buyers as well in such cooperation.

While in the AEU, buyers only have to define demand volumes for the delivery points (DP) of their choice, sellers also have to include a price in their bids in addition to quantity (which may be potentially different for each DP), thus sellers face a pricing problem under demand uncertainty, with the option of setting multiple prices simultaneously. A similar problem has been studied in the case of monopolies in \cite{dana2001monopoly}.

The non-binding nature of the AggregateEU (AEU) platform allows for overbidding (as the derived matches may be dropped later). Participants may be motivated to overbid either to (i) ensure a sufficient number of potential partners to match their desired demand/supply and/or to (ii) pursue bargain-hunting, that is, to cherry-pick the preferred cases from the determined matches.

Regarding the topic of non-binding mechanisms, the first pioneering works are focusing on the concept of \emph{cheap talk} \cite{farrell1996cheap}, where (non-binding) communication is restricted to the exchange of private information. In contrast, the concept of
indicative bidding assumes that non-binding bids submitted in the preliminary stage of the auction affect the resulting costs of bidders and their ability to bid again in the further stages of the auction process.
Indicative bidding and two-stage auctions in a one-sided single-item setting have been studied in
\cite{ye2007indicative}. The model used in this paper assumes (i) a preliminary stage, when the bidders are
asked to submit nonbinding bids, (ii) the restriction of the number of participants based on the prices indicated in the preliminary bids, (iii)  a
substantial entry cost for final bidders, and shows that there is no symmetric increasing (fully separating) equilibrium in this setting. As an alternative, the paper proposes to introduce auctions of entry rights to induce efficient entry. An analysis of the topic based on experimental data is described in \cite{kagel2008indicative}.
The more recent paper \cite{quint2018theory} assumes a similar setup, and characterizes the symmetric equilibria that may be identified in the setting.

\subsection{Objectives}

Although the AEU mechanism has a solid theoretical grounding \citep{Armstrong2006} it has also been accused of being an attempt to centralize decision-making on a European level. We are mainly interested in the effect of the non-binding nature. According to our best knowledge, no literature results are available about AEU-like non-binding allocation mechanisms for homogeneous goods, where the demand of buyers is aggregated, quantity-price offers are required from the side of sellers, and a price-ranking based matching is executed on multiple trading points simultaneously.


The objective of this paper is therefore to propose a quantitative modelling framework for describing the overbidding strategy in a non-binding demand aggregation/joint purchasing coordination platform with multiple delivery points (such as AEU) and to analyse its potential implications.

\section{The AggregateEU framework}
As of 2025, few academic contributions on the AggregateEU mechanism exist, as the platform is relatively new and limited information is available on its effects and operation. Most publications feature opinion pieces from traders and energy market experts, as well as summaries of the EC's decisions and news pieces on the results of auction rounds.

AggregateEU, operated by the PRISMA joint capacity booking platform, was launched in Q2 2023. PRISMA ceased operating the platform at the end of May 2025, from which date the European Commission will now undertake these activities on the EU Energy Platform.
Since its launch in Q2 2024, there have been five short-term and two mid-term tendering rounds. Short-term tenders delivered 43 bcm of matched gas, while the mid-term tenders matched 33.65 and 20 bcm of gas (see \citep{ACER2024}, \href{https://energy.ec.europa.eu/news/joint-gas-purchasing-very-good-results-2nd-mid-term-demand-aggregation-round-natural-gas-2025-03-26_en}{European Commission} and \href{https://www.rigzone.com/news/about_700_bcf_of_gas_matched_in_2nd_midterm_round_of_aggregateeu-27-mar-2025-180049-article/?utm_source=chatgpt.com}{Rigzone}).

\begin{table*}[t]
    \centering
\caption{Matched quantities under the {AggregateEU} mechanism}
\label{tab:AEU_data_1}
    \begin{tabular}{|c|r@{ - }l|r@{ - }l|c|c|c|}
   \hline
   \thead{Tender type\\ (term)}& \multicolumn{2}{c|}{\thead{Round period}} & \multicolumn{2}{c|}{\thead{Delivery period}} & \multicolumn{1}{c|}{\thead{Aggregated \\ demand \\ (bcm)}} & \multicolumn{1}{c|}{\thead{Supply\\ offers \\ (bcm)}} & \multicolumn{1}{c|}{\thead{Demand\\ matched \\ (bcm)}}  \\ \hline
         short&  Apr & May 23&  Jun 23 & May 24 &  11.6 &  18.7 &  10.9\\ \hline
         short&  Jun & Jul 23&  Aug 23 & Mar 25 &  15.9&  15.2 &  12\\ \hline
         short&  Sep & Oct 23 &  Dec 23 & Mar 25 &  18.1 &  16.5 &  11.9\\ \hline
         short&  Nov & Dec 23 &  Jan 24 & Mar 25&  9.1&  10.7 &  7.4\\ \hline
         short& Mar&Mar 24 &  May 24 & Mar 25 &  1.6&  2.01 &  1.12\\ \hline
         mid&  Feb&Feb 24&  Apr 24 & Sep 29 &  33.66 &  97.36&  33.65\\ \hline
          mid&  Mar&Mar 25&  Apr 25 & Sep 30 &  29 &  31&  20\\ \hline
 Total short& && && 56.3& 63.1&43.3\\ \hline
 Total mid& && && 63.66& 128.36&53.65\\ \hline
 Total & && && 119.96& 191.46&96.95\\
 \hline
 \end{tabular}
\end{table*}

The matched quantities are summarised in Table \ref{tab:AEU_data_1}.
There is unconfirmed anecdotal evidence that, as of September 2024, out of the 43 bcm of matched short-term contracts, only 1 bcm have been signed and reported to the Commission \href{}{\citep{hancock2024}}.

\cite{krajnik2024demand} provide a concise summary of the AggregateEU mechanism, detailing its main operation and the results of the first four short-term tenders as well as the first mid-term tender. By characterising the mechanism as a buyers’ group in economic literature, they provide a list of potential benefits, such as cost savings due to higher bargaining power, which can be passed on to consumers. At the same time, they also list potential inefficiencies, including the deadweight loss stemming from the monopsonic nature of a buyers' cartel as well as the so-called \textit{waterbed effect}. The waterbed effect refers to the phenomenon whereby the price discount gained by the buyers' cartel is the supplier's loss, subsequently recuperated from other buyers. The authors conclude that, due to a lack of data on auction results and concluded contracts, it is not possible to assess whether the price reduction for EU buyers occurred.

The International Gas Union 
\citep{IGU2024} notes that a joint purchasing scheme helps bolster supply security for smaller entities that pool demand, share infrastructure and risks, and collectively optimise purchasing costs. They refer to demand aggregation, a practice used by Japanese LNG buyers, and state that the AggregateEU platform strengthened the power of smaller European buyers to attract new sources to replace Russian natural gas, promote transparency, and reduce price volatility. Based on these examples, in 2023, KOGAS and JERA, major LNG importers in South Korea and Japan, agreed to expand cooperation in joint LNG purchasing and trading, including emergency swap agreements to strengthen energy security in both markets.

The views of traders' and energy company associations have generally been negative regarding the platform. A letter \citep{IOGP2023} on the prolongation of the joint purchasing argues that the platform is not transparent, and ``there is a lack of data justifying making it permanent.'' The signatories state that total matched volumes account for only 1 per cent of traded quantities at the reference trading hub, TTF.

The energy trader's association argues that ``market participants have been raising concerns that the mechanism, which is spearheaded by the European Commission, would cement the position of dominant players in certain EU countries'' \citep{sabadus2023}.   They also raise concerns regarding the confidential information of companies, which is relayed to the central buyer. In addition, they note that experienced traders are more likely to receive market rates and offers, whereas smaller, new entrants may pay higher prices. Overall, they claim that the platform had, at best, a neutral effect and would wither without further regulation.\footnote{``As long as people are not required to use it, it may die, or it could be held as a backstop in case there was another emergency scenario.''\citep{sabadus2023}}

Barnes \cite{barnes2022eu,barnes2023eu} criticises joint purchasing as i) the voluntary nature of taking part in demand aggregation makes it impractical. The total demand bid required makes up 15\% of the 90\% storage target for the winter of 2023--2024~\citep{eureg2022_1032}, which is around 13.5 bcm. This is negligible compared to the 400 bcm/year European demand and does not constitute a significant volume with buying power; ii) the mechanism creates tensions between Member States and the Commission by taking over national matters; and iii) an additional platform constitutes competition for the existing, well-developed energy exchanges.

The design of the mechanism is also being challenged: \cite{lecoq2023} argue that with non-binding matching, the buying power of aggregated demand is not realised at all. For smaller companies, the platform offers a central buyer or agent on behalf of, who acts as an intermediary. This is unfavourable for both smaller and larger companies: smaller ones are discouraged from sharing sensitive information, while larger ones are hindered by the weak bargaining power of smaller companies \citep{lecoq2023}.


In contrast, communication from the EC has been positive overall. Upon the closing of each tender round, the Commission delivered a press release on the matched quantities with a corresponding comment from the Vice-President Maro\v{s} Sef\v{c}ovi\v{c} on each round:\\
First round:
\begin{quote}
\it {``This is a remarkable success for an instrument that did not exist some five months ago. The Commission has played its role as an aggregator and matchmaker; now it is for the respective parties to conclude their agreements. It is a win-win for all parties.''} \citep{eu2023R1}
\end{quote}

\noindent Second round:
\begin{quote}
\it {``The EU Energy Platform on joint gas purchasing is working well and shows Europe at its best. Joining forces is key to guarantee stable and affordable gas supplies to the EU market, for the benefit of our businesses and citizens. As we enter a crucial stage of the storage refilling season, I encourage both buyers and suppliers to make the best use of this new marketplace and further stabilise the European gas market ahead of the winter.''} \citep{eu2023R2}
\end{quote}

\noindent Third round:
\begin{quote}
\it {``This year we are entering the winter with a much better outlook for our security of supply: our gas storage is nearly full, our supplies are more diverse and renewables are playing a more prominent role in our energy mix. However, the situation on the gas market remains tight, so we need to stay vigilant and continue purchasing gas together to guarantee stable and affordable supplies. I therefore invite European companies to once again join forces and work together to make sure we will get safely through this winter and already start preparing for the next one.''} \citep{eu2023R3}
\end{quote}

\noindent Fourth round:
\begin{quote}
\it {``This round of gas demand aggregation is another opportunity for European companies to join forces to purchase the energy they need. As we begin the heating season for most of Europe, the EU Energy Platform continues to contribute to our efforts to keep everybody warm this winter and keep energy at affordable prices, for citizens and European industry. The EU Energy Platform is a success story in terms of our collective response to last year’s energy crisis, and it continues to serve us today.''} \citep{eu2023R4}
\end{quote}

The Commission provides a brief evaluation of the mechanism in its country factsheet report \citep{eurepower2024}, including the volumes of aggregated demand for each Member State. Evidently, some Member States have greatly utilised the platform, aggregating over 25\% of their annual natural gas consumption (AT, BG, CZ, EE, FI, GR, HR, LU, SI, SK). Many of these countries were highly dependent on Russian natural gas imports at the outset and relied on the platform to secure alternative sources. Other Member States, where Russian natural gas represented only a small share of imports, made little use of the platform, with demand aggregated below 25\% (BE, CY, DK, ES, FR, IE, NL, MT, PT, SE).
Some major consumers of Russian natural gas chose to procure supplies by alternative means and did not aggregate a considerable volume of their demand on the platform (DE, IT, HU, LT, LV, PL, RO). Countries also differ from the point of view of liquidity on the wholasale gas market including access to alternative sources. Furthermore, they also diverge on gas demand flexibility, affected by sectoral demand and fuel switching options - these are reflected in their aggregated demand. Some countries have a relatively small share of natural gas in their primary energy mix, therefore they placed small or no bids (DK and SE). Market liquidity and potential to access alternative sources also plays a great role: those countries with illiquid markets and limited alternatives may place higher bids (e.g. FI, EE), while countries having mature gas markets and more alternatives may have chosen not ot utilise the platform (e.g. FR, BE, ES, NL). (See Table \ref{tab:AEU_data_2})

In May 2024, the Commission launched a public consultation on the potential extension of the mechanism for long-term tenders (lasting over 20 years). Unfortunately, the results of the consultation are not publicly available. Overall, the AggregateEU mechanism has received mixed opinions. Whereas the Commission is mostly positive (but does not back up its evaluation with any data), the major traders and trading associations are more sceptical.

\begin{table}
    \centering
    \begin{tabular}{|c|c|c|c|} \hline
         Country&  \multicolumn{1}{c|}{\thead{Aggregate\\  demand}}& \multicolumn{1}{c|}{\thead{Share in  annual\\ gas consumption}}&\multicolumn{1}{c|}{\thead{Annual gas\\ consumption (bcm)*}}\\ \hline
         AT&  2.35& 28.70\%&8.2\\ \hline
         BE&  0.41& 2.63\%&15.6\\ \hline
         BG&  0.81& 29.86\%&2.7\\ \hline
         CY&  0& 0.00\%&0.0\\ \hline
         CZ&  11.95& 156.96\%&7.6\\ \hline
         DE&  4.13& 4.61\%&89.6\\ \hline
         DK&  0& 0.00\%&0.0\\ \hline
         EE&  0.22& 60.54\%&0.4\\ \hline
         ES&  4.28& 13.12\%&32.6\\ \hline
 FI& 0.45& 33.63\%&1.3\\\hline
 FR& 1.37& 3.63\%&37.7\\\hline
 GR& 6.15& 118.90\%&5.2\\\hline
 HR& 0.76& 30.04\%&2.5\\\hline
 HU& 0.64& 6.59\%&9.7\\\hline
 IE& 0.08& 1.51\%&5.3\\\hline
 IT& 8.32& 12.11\%&68.7\\\hline
 LT& 0.09& 5.87\%&1.5\\\hline
 LU& 0.22& 37.13\%&0.6\\\hline
 LV& 0.09& 10.67\%&0.8\\\hline
 MT& 0& 0.00\%&0.0\\\hline
 NL& 5.74& 17.18\%&33.4\\\hline
 PL& 1.29& 6.55\%&19.7\\\hline
 PT& 0.04& 0.72\%&5.6\\\hline
 RO& 0.14& 1.40\%&10.0\\\hline
 SE& 0& 0.00\%&0.0\\\hline
 SI& 0.25& 29.53\%&0.8\\\hline
 SK& 2.35& 52.00\%&4.5\\\hline
 Total& 52.13& 14.32\%&364.00\\\hline
    \end{tabular}
    \caption{Aggregated demand per Member State \citep{eurepower2024}\\\textit{*`Annual gas consumption' was calculated as the ratio of `Aggregated demand' and `Share in annual gas consumption'}}
    \label{tab:AEU_data_2}
\end{table}

In 2024, Frontier Economics prepared an analysis of the first 4 rounds of AggregateEU
\citep{doi/10.2833/4362329}. In the first four rounds, 48 sellers participated, with eight of them trading with LNG. Concentration among the eight sellers was high, with 67\% of the total LNG offered coming from a single seller. Matches totalled 42.2~bcm for the four short-term rounds in total, of which 28.45 bcm was pipeline gas and 13.67 bcm was LNG. Matched demand was higher in more liquid Western European markets, while less liquid Eastern European countries reported lower demand and a lower matching rate.
Throughout the tenders, the share of unmatched demand increased from 18\% to 37\%.

Regarding the effect of AggregateEU on the market, the report \citep{doi/10.2833/4362329} states that the new platform \emph{'has likely been a contribution to establishing new commercial relationships. For example, based on voluntary reporting by AggregateEU participants, at least 9
buyers were matched with between 1 and 5 counterparties each, with whom they did not have
prior contractual relationships.'}. An other statement of the report is that \emph{'it is plausible that, by signalling the EU's willingness to act collectively on global markets when faced with an external threat,
AggregateEU may help reduce the likelihood of third countries seeking to use trade to exert
pressure on the EU.'}. Furthermore, while the report highlights that LNG volumes procured through AggregateEU were low as a share of total
demand, it suggests that \emph{'it is possible that participation in AggregateEU may have generated positive spillovers in terms of LNG procurement in the wider market, for example in terms of establishing new relationships with LNG suppliers to the EU.'} (although the report admits that there is no evidence for this having been the case). Regarding the prices, the report formulates the statement that \emph{'The limited stakeholder feedback on this subject (see below) suggests that, overall, prices
reached via AggregateEU are unlikely to have been vastly different to those of the wider
market.'}.

In April 2025, PRISMA announced that in the future, \emph{'the European Commission is responsible for operating AggregateEU to ensure continuity of the demand aggregation and joint purchasing of gas while developing a permanent gas mechanism.'}

\begin{table}[h!]
    \centering
\caption{LNG bids and matches of the first 4 rounds on {AggregateEU}, in bcm \citep{doi/10.2833/4362329}}
\label{tab:AEU_data_3}
    \begin{tabular}{|c|c|c|c|c|}
   \hline
   {Round}& {Matched} & {Unmatched} & {Total} &{Unmatched}  \\ \hline
         1&  2.21& 0.49  &  2.70  & 18\% \\ \hline
         2&5.56 & 2.18  &  7.74    & 28\% \\ \hline
         3&  4.32 &  2.04 &  6.36  & 32\%\\ \hline
         4& 1.58  & 0.92 & 2.50    & 37\%\\ \hline
         Total&13.67&5.63&19.3&29\%\\ \hline

    \end{tabular}
\end{table}

Given the limited data availability, analyses and evaluations of the platform's functioning and efficiency are currently lacking. Since we do not have access to auction results, our paper contributes to the ex-post evaluation of the policy by creating a stylised model that represents how the platform works. Based on the general observations made from the model and the formulation of the platform, we can identify the main advantages and disadvantages of the existing mechanism and potentially suggest adjustments.

\subsection{Aspects of the non-binding nature}

The platform matches the consumers to suppliers, among whom a potential deal seems realistic, and leaves the rest to the participants. As the matching is non-binding, any participant may withdraw from a potential deal at any time.
As the EC often uses the expression `the Tinder of the Gas market' for the AggregateEU platform to emphasise its non-binding nature, we begin by showing our key ideas in a simple Tinder model.

\begin{ex}
This example illustrates that overbidding on a matching platform may help establish connections between supply and demand, thereby creating matches.

We consider a marriage problem \citep{gale1962college} with non-complete preference lists, but we do not use the deferred acceptance (Gale-Shapley) algorithm; instead, we employ a one-round `rapid matching' approach. Men give one or (in the case of overbidding) more offers according to their preference list, and women respond according to their preference list: If a woman has multiple acceptable offers, she accepts the most preferred one (or in the case of overbidding, the most preferred ones). If a woman has only one acceptable offer, she accepts that. If there are no offers from acceptable partners, all offers are rejected.

Let us take 3 men ($A$, $B$, and $C$) and 3 women ($D$, $E$, $F$) with the following incomplete preference lists (Elements not present in the list are not acceptable.):
\begin{align*}
& A: F \succ E \succ D\quad\quad && D: C \succ A\\
& B: D \succ F \succ E\quad\quad && E: A \succ B \\
& C: E \succ D \succ F\quad\quad && F: B \succ C
\end{align*}
Everyone is up to only 1 date.

\paragraph*{Case 1 (No overbidding)}
No overbidding implies a single offer/acceptance response per person. We have the following offers:
\begin{align*}
A \rightarrow F, ~~B \rightarrow D, ~~C \rightarrow E
\end{align*}
No woman receives an acceptable offer, no matches, no dates.

\paragraph*{Case 2 (Overbidding)}
Under overbidding, we allow multiple (here: two) offers/acceptance replies per person. We have the following offers:
\begin{align*}
& A \rightarrow F, ~~B \rightarrow D, ~~C \rightarrow E \\
& A \rightarrow E, ~~B \rightarrow F, ~~C \rightarrow D
\end{align*}
Each woman receives one acceptable offer, and the following matches (dates) are established: $A-E, B-F$, and $C-D$.
\end{ex}

\begin{ex}
In the previous example, as no participant was \textit{overmatched} (matched with more than one partner), no matches were dropped.
On the other hand, as illustrated here, more matches due to the overbidding of participants may also result in fewer dates (realised matches) if the configuration of how less preferred matches are unilaterally dropped is unfortunate.
\begin{align*}
& A: D \succ E \succ F && D: C \succ A\\
& B: E \succ F \succ D && E: A \succ B\\
& C: F \succ D \succ E && F: B \succ C
\end{align*}

\paragraph*{Case 1 (No overbidding)}
We have the following offers:
\begin{align*}
A \rightarrow D, ~~B \rightarrow E, ~~C \rightarrow F
\end{align*}
As each woman receives only one offer, which is acceptable (no better alternatives are present), all offers are accepted, forming three matches. Each participant has exactly one match, no alternatives, and all matches are realised, resulting in 3 dates.

\paragraph*{Case 2 (Overbidding)}
In this case, the offers are as follows.
\begin{align*}
& A \rightarrow D, ~~B \rightarrow E, ~~C \rightarrow F \\
& A \rightarrow E, ~~B \rightarrow F, ~~C \rightarrow D
\end{align*}

As two acceptance replies are allowed for women, and each incoming offer of each woman is from an acceptable man, all offers are accepted, resulting in six matches. Meanwhile, every participant is up to only one date, thus each participant drops the least preferred match. We assume that this happens simultaneously, such that no participant is informed whether he/she have been dropped in a match before he/she make his/her decision regarding the drop of matches.

\begin{itemize}
    \item $A$ has 2 matches ($D$ and $E$), prefers $D$, so he drops $E$.
    \item $B$ has 2 matches ($E$ and $F$), prefers $E$, so he drops $F$.
    \item $C$ has 2 matches ($D$ and $F$), prefers $F$, so he drops $D$.
    \item $D$ has 2 matches ($A$ and $C$), prefers $C$, so she drops $A$.
    \item $E$ has 2 matches ($A$ and $B$), prefers $A$, so she drops $B$.
    \item $F$ has 2 matches ($B$ and $C$), prefers $B$, so she drops $C$.
\end{itemize}

As we can see, all matches are cancelled, and no dates are realised.
\end{ex}

\paragraph{Interpretation}
How can we interpret the above `Tinder' examples in the framework of the Aggregate EU mechanism (AEU)? If we assume that every participant in the AEU may be characterised with a fixed quantity describing the available or required amount of natural gas, this amount corresponds to the assumption that the participant in question is up only to one (or a fixed number of) date(s), independent of the potential number of matches. Overbidding in the AEU (that is, submitting bids with an overall quantity exceeding the available/required quantity) corresponds to the `swipe a lot to the right, and we'll see' strategy: participating with a quota exceeding the number of possible realised matches in which the participant may eventually be involved. In such a case, overmatching implies that the participant will drop the least preferred matches.

Certain differences, however, limit the validity of this analogy. Delivery points (DPs) represent a third, intermediate party in the matching process within the AEU. Consumers and suppliers have preferences over DPs (and not over each other), based on, for example, transfer costs, and determine their bids accordingly. Multiple DPs may be interpreted as a `simultaneous presence' in multiple matching platforms, each with limited accessibility to participants (as not every DP is available for each participant). Furthermore, while items (individuals) in the Tinder example/marriage problem are indivisible, natural gas is a divisible good.

For the AEU, only the matched but not the contracted quantities are reported, and therefore, it is impossible to estimate the ratio of potential overbidding and dropped matches.

In the following, a simplified quantitative model is proposed, which allows the analysis of the phenomenon and the implications of overbidding in the AEU. With the use of this model, we show that similar to the previously discussed `Tinder' examples, overbidding in the AEU has unpredictable implications, too, depending on the parameters of the actual scenario. Additionally, we provide some insights into the strategic aspects of the implied game and its equilibrium properties.

\section{Model}
\label{sec:model}
We aim to create a simple mathematical model of the AEU setup that can provide a quantitative description of the strategic decision of a participant overbidding or not. In the proposed model, the prior defined parameters and the strategy choices of participants explicitly define the model variables: the submitted bids, from which the matched quantities are derived via the pro-rata mechanism at each delivery point. As we will see, the contracted quantities are also explicitly determined from the matched quantities, according to simple assumptions (i.e., by the simultaneous drop of the least preferred bids by each player).

Accordingly, we model the interaction among buyers, sellers, and delivery points using the following strategic game.
We consider $n_C$ consumers (buyers) $n_S$ suppliers (sellers), and $n_{\DP}$ trading/delivery points (DPs)
In general, we consider both consumers and suppliers strategic actors and, therefore, players of the game. However, the proposed framework may be easily modified to study the interactions of players on one side, where only consumers or suppliers are strategic, and the decisions of the other side are given.

\subsection{Model parameters}

\begin{table*}[t]
    \centering
    \begin{tabular}{|c|c|}
    \hline
         Notation & Parameter\\
         \hline
        $n_C$ &number of consumers  \\$n_S$ & number of suppliers \\
         $qr^C_i$ &required quantity of comsumer $i$ \\
          $u^C_i$  &  utility of comsumer $i$ \\
          $\overline{q}^C_{i,t}$ &  maximal transferable quantity of consumer $i$ from DP $t$ \\
            $ct^C_{i,t}$  & unit transfer cost of consumer $i$ regarding DP $t$\\
           $qa^S_j$  & available quantity of supplier $j$\\
          $cp^S_j$  & unit production cost of supplier $j$ \\
          $\overline{q}^S_{j,t}$ &  maximal transferable quantity of supplier $j$ to DP $t$ \\
          $ct^S_{j,t}$  & unit transfer cost of supplier $j$ regarding DP $t$ \\
         \hline
    \end{tabular}
    \caption{Model parameters}
    \label{Tab_parameters}
\end{table*}

Model parameters are summarised in Table \ref{Tab_parameters}.
The parameters of consumers and suppliers are denoted with the superscripts $^C$ and $^S$, respectively. Consumers, suppliers and DPs are indexed by $i \in \{1,..n_C\}$, $j \in \{1,..n_S\}$ and $t \in \{1,..n_{\DP}\}$ respectively.
Each consumer $i\in \{1, ...,~n_C\}$ is characterized by the required quantity $qr^C_i$, the utility $u^C_i$, which describes the amount he is willing to pay for one unit of gas, and the vectors  $\overline{q}^C_{i},~ct^C_{i}\in\Real^{n_{\DP}}$, describing the access constraints and \textit{transfer costs} corresponding to the various DPs respectively:
$\overline{q}^C_{i,t}$ is the maximal quantity of consumer $i$ can physically transfer from DP $t$ --- possibly 0 ---; and $ct^C_{i,t}$ is the unit transfer cost consumer $i$ pays if he buys gas at DP $t$.

Likewise, each supplier $j\in \{1, ...,~n_S\}$ is characterised by the available quantity $qa^S_j$, the unit production cost $cp^S_j$, and the vectors  $\overline{q}^S_{j},~ct^C_{j}\in\Real^{n_{\DP}}$, describing the access constraints and \textit{transfer costs} corresponding to the various DPs
$\overline{q}^S_{j,t}$ is the maximal quantity supplier $j$ can physically transfer to DP $t$ -- again, possibly 0; and $ct^S_{j,t}$ is the unit transfer cost paid by supplier $j$ when selling gas at DP $t$.

\subsection{Formal description of the submitted bids}
\label{subsec:bid_format}

As mentioned before, the submitted bids, determined by the model parameters and the strategy choices of participants, may be regarded as variables of the model. In this subsection, we introduce the respective notations and describe how the bids are derived in the proposed model.

As consumers submit only demand quantities for DPs in the AEU platform, the bids of consumers in the model are summarised by the matrix

    $$
    BQ^{C}\in\Real_+^{n_C\times n_{\DP}}~~~,
    $$
    where the entry $BQ^C(i,t)$ corresponds to the quantity of the demand (buy) bid submitted by consumer $i$ for DP $t$.
    We assume that each demand bid submitted to a particular DP is bounded by the accessibility constraint and the total required quantity of the respective consumer, as described by equation (\ref{eq:upper_bounds_for_BQC(i,t)}).

    \begin{equation}
    \label{eq:upper_bounds_for_BQC(i,t)}
        BQ^C(i,t) \leq \overline{q}^C_{i,t}~~~~BQ^C(i,t) \leq qr^C_i ~~~\forall i,~\forall t
    \end{equation}

    Suppliers also have to provide bid prices; thus, the bids of suppliers are summarised in the matrices

    $$
    BQ^{S}\in\Real_+^{n_S\times n_{\DP}}~~~\text{and}~~~BP^{S}\in\Real_+^{n_S\times n_{\DP}}~~~,
    $$

    where $BQ^S(j,t)$ corresponds to the quantity of the supply (sell) bid submitted by supplier $j$ for DP $t$ and
    $BP^S(j,t)$ corresponds to the price of the bid submitted by supplier $j$ for DP $t$.

    Similarly to the demand side, we assume that each supply bid submitted to a particular DP is bounded by the accessibility constraint and the total required quantity of the respective supplier, as described by equation (\ref{eq:upper_bounds_for_BQS(j,t)}).

    \begin{equation}
    \label{eq:upper_bounds_for_BQS(j,t)}
        BQ^S(j,t) \leq \overline{q}^S_{j,t}~~~~BQ^S(j,t) \leq qa^S_j~~~\forall j,~\forall t
    \end{equation}

   The total quantity of submitted demand and supply bids for DP $t$ is denoted by $q^{D}_t$ and $q^{S}_t$ and may be calculated as described by Equation~\eqref{eq_qDt_qSt}.

\begin{equation}\label{eq_qDt_qSt}
    q^{D}_t=\sum_{i} BQ^C(i,t),~~~~~~~~~~ q^{S}_t=\sum_{j} BQ^S(j,t)
\end{equation}

\subsection{Bidding strategies}
\label{subsec:bidding_strat}

We model the optional overbidding as a strategic decision. Each participant may choose from two options: (1) Not overbid, or (2) Overbid.
$S_i\in \{ N,O\}$ denotes the strategy of player $i$, where $N$ stands for non-overbidding and $O$ stands for overbidding.

In the former case ('N'), the total quantity of the bids submitted by the participant will be equal to the required/available quantity (in the case of consumers and suppliers, respectively). If $\overline{q}^C_i$/$\overline{q}^S_i$ limits the total quantity of bids, the participant will submit bids for the maximal total (deliverable) quantity.
Formally, the non-overbidding strategy results in a total demanded/supplied quantity (denoted by $TDQ^N_i$ and $TSQ^N_j$ respectively)

\begin{align*}
 &   TDQ^N_i=\min\set{qr^{C}_i,\sum \overline{q}^C_{i}}
    \text{ for consumers, and}\\
 &    TSQ^N_j=\min\set{qa^{S}_j,\sum \overline{q}^S_{j}} \text{ for suppliers}
\end{align*}

In the latter case ('O'), if the $\overline{q}^C_i$/$\overline{q}^S_i$ constraints allow, the total bidded quantity will be equal to two times the required/available quantity. Again,  if the accessibility constraint limits the total quantity of bids, the participant will submit bids for the maximal total (deliverable) quantity. In addition, according to inequalities (\ref{eq:upper_bounds_for_BQC(i,t)}) and (\ref{eq:upper_bounds_for_BQS(j,t)}) we assume that no consumer $i$ or supplier $j$ submits a bid greater than $qr^C_i$ or $qa^S_j$, respectively, to any single DP, even if the the accessibility constraint would allow it, in the case of overbidding either. Formally, the overbidding strategy results in a total demanded/supplied quantity (denoted by $TDQ^O_i$ and $TSQ^O_j$ respectively)

\begin{align*}
 &   TDQ^O_i=\min\set{2qr^{C}_i,\sum \overline{q}^C_{i}}
    \text{ for consumers, and}\\
 &    TSQ^O_j=\min\set{2qa^{S}_j,\sum \overline{q}^S_{j}} \text{ for suppliers}
\end{align*}


\subsubsection{Bidding strategy of consumers}
\label{subsec:bidding_cons}
Irrespective of the overall bid quantities, the bidding protocol requires a consumer to specify the allocation of this total across delivery points.

We assume a straightforward bidding strategy for consumers. Once the total quantity of the submitted bids ($TDQ_i=TDQ^N_i$ or $TDQ_i=TDQ^O_i$) is determined by the actual strategy $S_i\in\set{N, O}$, we assume that consumer $i$ allocates this quantity, minimising the total access cost. Formally, consumer $i$ solves a linear programming (LP) problem described by equations (\ref{eq_cons_LP_obj} -- \ref{eq_cons_LP_constr}), where $x  \in\Real^{n_{\DP}}$.

\begin{align}
& \min_x \quad \sum_{t \in \{1,...,n_{DP}\}} x_t~ ct^C_{i,t}  \quad st. \label{eq_cons_LP_obj} \\
& \sum_t x_t = TDQ_i \quad x_t \leq \overline{q}^C_{i,t}~~\forall t, \quad x_t \leq qr^C_i ~~\forall t \label{eq_cons_LP_constr}
\end{align}

The result of the LP will determine the bids of consumer $i$, i.e. the $i$-th row of the matrix $BQ^C$ (denoted by $BQ^C(i,.)$) as

$$
BQ^C(i,.)=x^T \quad .
$$

\begin{ex} Let us consider a single consumer and 3 DPs, described by the parameters $qr^C_1= 10$, $\overline{q}^C_1= [6~~ 8~~ 9]$, $ct^C_1=[3~~2.5~~2]$. Since $ct^C_1$ shows that DP 3 has the lowest access cost, followed by DP 2 and DP 1, in the case of the strategy $S_1=N$, the submitted bids will be described by
\begin{align*}
    BQ^{C,N}(1,.)=[0 ~~1~~9] ~~~.
\end{align*}

In the case of overbidding ($S_1=O$), the submitted bids will be
\begin{align*}
    BQ^{C,O}(1,.)=[3 ~~8~~9]~~~.
\end{align*}
\end{ex}

\subsubsection{Bidding strategy of suppliers}
\label{subsec:bidding_sup}

Likewise, suppliers distribute the total bid quantity (equal to $TSQ^N_j$ or $TSQ^O_j$, depending on $S_j$) among DPs, according to the same principle as consumers: DPs with lower access costs are preferred. The respective LP may be derived similarly. The solution of the LP of supplier $j$ determines $BQ^S(j,.)$.
However, in the case of suppliers, a bid price has to be determined for each bid.

\paragraph{Pricing model:} In the current work, we assume a very simple pricing mechanism as follows. Each supplier $j$ determines a baseline price, denoted by $p^{\BL}_j$, and a transfer/cost dependent additive component $pt^{S}_{j,t}$ for each DP $t$. According to our assumptions, the baseline price is equal to the lowest price, which ensures non-negative profit over all possible DPs, considering the costs of production and transfer, that is,

$$
p^{\BL}_j=cp^S_j+\max_t ct^S_{j,t}~~~.
$$

Regarding the additive component, we assume that it is equal to half of the respective transfer cost.

$$
p^{T}_{j,t}=\frac{1}{2} ct^S_{j,t}
$$

The resulting bid price of supplier $j$ for DP $t$ ($BP^S(j,t)$) may be calculated as described by Equation~\eqref{eq:p_B_ij}.

\begin{equation}
    BP^S(j,t)=p^{\BL}_j + p^{T}_{j,t} \label{eq:p_B_ij}
\end{equation}

While the formula described by Equation~(\ref{eq:p_B_ij}) ensures positive profit at any DP for the supplier in question, it also reflects the different access costs of DPs in the bid prices.

\subsection{Modelling the operation of the coordination platform and contracting}
\label{subsec:model_coord_and_contracting}

We study how the submitted bids influence the matches returned by the coordination platform and how these matches ultimately determine the final contracts.
As in the case of the AggregateEU, the matches determined by the coordination platform are non-binding; thus, participants may cancel (or `drop') a deal before contracting without further ado. In our model, each participant selects the best deals up to their desired or available quantity and discards the rest of the alternatives.
For simplicity, we consider a single-period market consisting of three stages and assume that each participant may submit only one bid for each DP. In the following, we elaborate on these stages.
\begin{description}
    \item[1. Bidding] Participants choose their strategy ('N' or 'O') and submit their bids ($BQ^C$, $BQ^S$, and $BP^S$) to the coordination platform, according to the considerations described in subsection \ref{subsec:bidding_strat}. The submitted bids determine the $q^D_t$ and $q^S_t$ values for the individual DPs as described by eq. \eqref{eq_qDt_qSt}.

    \item[2. Matching] The coordination platform determines the matching of the submitted bids, according to the principles of the AggregateEU framework: (i) if there is oversupply in a DP, cheaper supply bids are given priority and (ii) to ensure fair access to cheap sources, all matched supply bids are distributed among consumers `\textit{pro rata}', that is, proportional to the quantity of their submitted demand bids. The matching process may be described in detail as follows.

    In every DP, the total quantity of submitted demand and supply bids is determined. If the total demand is more than or equal to the total supply in the actual DP $t$ (that is, $q^{D}_t>q^{S}_t$), the supply bids are matched to the demand bids \textit{pro rata}: each supply bid is divided into $n^C_t$ parts, where $n^C_t=\abs{BQ^C(.,t)>0}$ denotes the number of consumers, who submitted demand bids (with positive quantity) for the actual DP $t$. The volume of the parts is proportional to the volume of the submitted demand bids. If the total quantity of the submitted supply bids exceeds the total quantity of the submitted demand bids (that is, $q^{D}_t<q^{S}_t$), the most expensive supply bids are fully or partially dropped, until the total quantities match, and the \textit{pro rata} matching with demand bids is applied to the remaining supply bids. We denote a match by
    $(C_i-S_j, t)$, where $i$ and $j$ are the indices of the matched consumer and supplier, while $t$ is the index of the DP where the match is established.

    \item[3. Contracting] After the matches have been determined, it may happen for particular overbidding consumers (suppliers) that the matched or allocated quantity will be higher compared to the quantity required by the consumer (available by the supplier, respectively). In this case, as the mechanism is non-binding, it is straightforward to assume that the player in question drops all but the most preferred deals up to its capacity limit. If multiple matches are present for a participant implying the same utility, we assume they will be dropped at the same rate (partially or fully).
    According to the modelling assumption, matches, which are not dropped by either concerned participants (the respective consumer and supplier) are realized as contracts with the price determined by the matching.
    \item[4. Utilities] Now the resulting utility values of consumers and suppliers may be calculated according to equations
(\ref{eq:U^C}) and (\ref{eq:U^S}).
\begin{align}
    U^C_i&= Q^C_i u^C_i - C^C_i \label{eq:U^C} \\
    U^S_j&= I^S_j - C^S_j, \label{eq:U^S}
\end{align}
where $Q^C_i$ denotes the total amount of gas received by consumer $i$ according to the realised contracts, $I^S_j$ denotes the total income of supplier $j$ according to the realised contracts, while $C^C_i$ stands for the total cost of consumer $i$, i.e. the sum of expenses of the realised bargains and transfer costs, and
$C^S_j$ denotes the total cost of supplier $j$, composed of production and transfer costs.
\end{description}

\section{Results}\label{sec:results}

\subsection{The ambiguous effect of overbidding}
\label{subsec:results_1st_subsec}

First, similar to the Tinder examples discussed in the introduction, we limit our analysis to cases where all players follow the same strategy. They either truthfully bid their required/available quantities, or they all overbid and bid twice the required/available amount (subject to access constraints in both cases, but these will not be binding in our examples).

To illustrate how uncertain the potential effect of overbidding on the contracted outcomes may be, we use two simple examples (scenarios).
While the first scenario demonstrates the potential benefits of overbidding, the second scenario illustrates its possible drawbacks.


\begin{ex}[Scenario I]
\label{subsec:Ex1}

This scenario illustrates that overbidding may help to connect demand with supply in cases where they wouldn't be matched otherwise. This can be illustrated by a very simple example involving a single consumer (C), a single supplier (S), both with unitary demand and supply quantities, and three delivery points (DPs). Let us assume that the access limit of both C and S to each of the DPs is 1 unit, and transfer costs are such that the preference list of C is $DP_1 \succ DP_2 \succ DP_3$, while the preference list of S implies $DP_3 \succ DP_2 \succ DP_1$. The parameters are summarized in equations (\ref{eq_par_cons}) and (\ref{eq_par_sup}).

\begin{align}
     qr^C_1&= 1 \quad &u^C_1&=2
     \nonumber \\\overline{q}^C_1&= [1~~ 1~~1] \quad &ct^C_1&=[0.1~~ 0.2~~0.3]
     \label{eq_par_cons} \\
    qa^S_1&= 1 \quad &cp^S_1&=1 \nonumber \\ \overline{q}^S_1&= [1~~ 1~~1] \quad &ct^S_1&=[0.3~~ 0.2~~ 0.1] \label{eq_par_sup}
\end{align}

The submitted bids may be calculated according to subsections \ref{subsec:bidding_cons} and \ref{subsec:bidding_sup}, based on the applied strategy of participants -- non-overbidding (upper index $N$) or overbidding (upper index $O$) and on the parameter values described in Equations \eqref{eq_par_cons} and \eqref{eq_par_sup}, and they will be as summarised in Figs. \ref{fig:AEU_new_E1_NO} and \ref{fig:AEU_new_E1_O}  and described by \eqref{eq:BQCN_BQSN_Ex1}, \eqref{eq:BQCO_BQSO_Ex1} and \eqref{eq:BPSN_BPSO_Ex1}.

\begin{figure}[h!]
    \centering
    \includegraphics[width=0.8\linewidth]{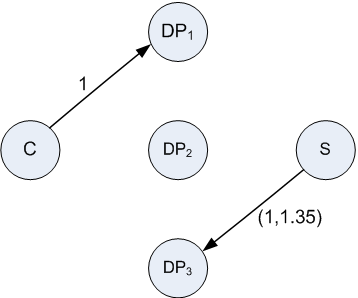}
    \caption{Bids of Scenario I in the case of non-overbidding (N). In the case of demand bids, the bidded quantity, in the case of supply bids, the bidded quantity and the respective price are indicated.}
    \label{fig:AEU_new_E1_NO}
\end{figure}

\begin{figure}[h!]
    \centering
    \includegraphics[width=0.8\linewidth]{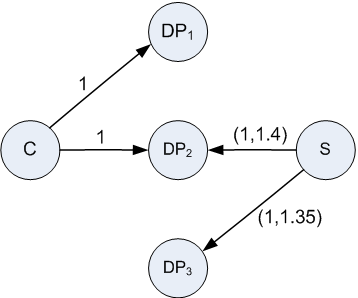}
    \caption{Bids of Scenario I in the case of overbidding (O).}
    \label{fig:AEU_new_E1_O}
\end{figure}

\begin{small}
\begin{align}
    BQ^{C,N}=\left( \begin{array}{ccc}
         1 & 0 & 0
    \end{array} \right)\nonumber \\
    BQ^{S,N}=\left( \begin{array}{ccc}
         0 & 0 & 1
    \end{array} \right) \label{eq:BQCN_BQSN_Ex1}
\end{align}
\begin{align}
    BQ^{C,O}=\left( \begin{array}{ccc}
         1 & 1 & 0
    \end{array} \right)\nonumber \\
    BQ^{S,O}=\left( \begin{array}{ccc}
         0 & 1 & 1
    \end{array} \right) \label{eq:BQCO_BQSO_Ex1}
    \end{align}
 \begin{equation}   \label{eq:BPSN_BPSO_Ex1}
BP^{S,N}=BP^{S,O}=\left( \begin{array}{ccc}
         1.45 & 1.4 & 1.35
    \end{array} \right)
\end{equation}
\end{small}


Regarding the non-overbidding (N) case, as fig.\ref{fig:AEU_new_E1_NO} shows, the supply does not meet the demand, no matches are established, and the resulting total matched quantity ($TMQ$) (thus also the total contracted quantity ($TCQ$)), and the resulting utility of players is 0.

In contrast, in the case of overbidding (O), as fig. \ref{fig:AEU_new_E1_O} shows, each participant bids its required/available quantity both for its most preferred and its second most preferred DP (DP$_1$ and DP$_2$ in the case of C, and DP$_3$ and DP$_2$ in the case of S), thus a match is established at DP$_2$. As this match is the only one both for C and S, they do not drop it, thus it will be a fully contracted match, resulting in $TMQ=TCQ=1$, implying a resulting utility of 0.4 for C and 0.2 for S.

Overall, in this particular case, uniform overbidding increases the total contracted quantity (TCQ) and total resulting utility (TU) compared to the non-overbidding case (1 and 0.6 vs 0 and 0, respectively).
\end{ex}


\begin{ex}[Scenario II]
\label{subsec:Ex2}

Scenario II illustrates that overbidding may also be a source of inefficiency, through overmatching and the asymmetric drop of matches. We consider two consumers (C$_1$, C$_2$) and two suppliers (S$_1$, S$_2$).
While the consumers are both characterised by 1 unit of demand, the available quantity of S$_1$ ($qa^S_1$) equals 2 units, while
S$_2$, which is assumed to be the cheaper source ($cp^S_1=2$, while $cp^S_2=1$), offers 1 unit of gas ($qa^S_2=1$).
We assume that the access limit of both C$_1$ and C$_2$ to each of the DPs is 1 unit, the access limit of S$_1$ to each of the DPs is 2 units, while the access limit of S$_2$ to each of the DPs is 1 unit.
Transfer costs are such that the preference list of C$_1$ is $DP_2 \succ DP_1 \succ DP_3$, the preference list of C$_2$ is $DP_2 \succ DP_3 \succ DP_1$, the preference list of S$_1$ is $DP_2 \succ DP_1 \succ DP_3$ and
the preference list of S$_2$ is $DP_3 \succ DP_1 \succ DP_2$. The parameters are summarised in equations (\ref{eq_par_cons_2}) and (\ref{eq_par_sup_2}).

\begin{align}
     qr^C_1&= 1 \quad &u^C_1&=5
     \nonumber \\\overline{q}^C_1&= [1~~ 1~~1] \quad &ct^C_1&=[0.2~~ 0.1~~0.3] \nonumber \\
     qr^C_2&= 1 \quad &u^C_2&=5
     \nonumber \\\overline{q}^C_2&= [1~~ 1~~1] \quad &ct^C_2&=[0.3~~ 0.1~~0.2]
     \label{eq_par_cons_2} \\
    qa^S_1&= 2 \quad &cp^S_1&=2 \nonumber \\ \overline{q}^S_1&= [2~~ 2~~2] \quad &ct^S_1&=[0.3~~ 0.2~~ 0.4]  \nonumber \\
    qa^S_2&= 1 \quad &cp^S_2&=1 \nonumber \\ \overline{q}^S_2&= [1~~ 1~~1] \quad &ct^S_2&=[0.3~~ 0.4~~ 0.2]\label{eq_par_sup_2}
\end{align}

The submitted bids in the case of non-overbidding (N) and overbidding (O) are calculated similarly to Scenario I, and they are summarised in Figs. \ref{fig:AEU_new_E2_NO} and \ref{fig:AEU_new_E2_O}  and described by \eqref{eq:BQCN_BQSN_Ex2}, \eqref{eq:BQCO_BQSO_Ex2} and \eqref{eq:BPSN_BPSO_Ex2}.

\begin{figure}[h!]
    \centering
    \includegraphics[width=0.8\linewidth]{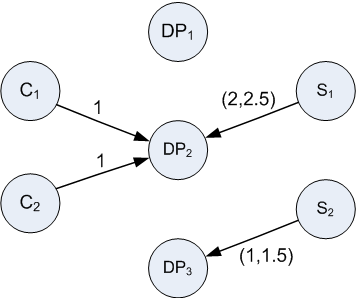}
    \caption{Bids of Scenario II in the case of non-overbidding (N). }
        \label{fig:AEU_new_E2_NO}
\end{figure}

\begin{figure}[h!]
    \centering
    \includegraphics[width=0.8\linewidth]{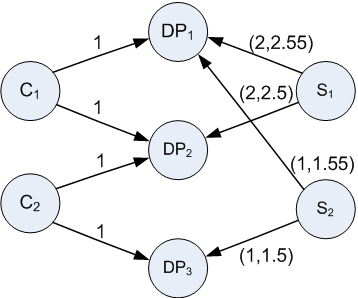}
    \caption{Bids of Scenario II in the case of overbidding (O).}
    \label{fig:AEU_new_E2_O}
\end{figure}

\begin{small}
\begin{align}
    BQ^{C,N}=\left( \begin{array}{ccc}
         0 & 1 & 0 \\
         0 & 1 & 0 \\
    \end{array} \right)\nonumber \\
    BQ^{S,N}=\left( \begin{array}{ccc}
         0 & 2 & 0 \\
         0 & 0 & 1
    \end{array} \right) \label{eq:BQCN_BQSN_Ex2}
\end{align}
\begin{align}
    BQ^{C,O}=\left( \begin{array}{ccc}
         1 & 1 & 0 \\
         0 & 1 & 1
    \end{array} \right)\nonumber \\
    BQ^{S,O}=\left( \begin{array}{ccc}
         2 & 2 & 0 \\
         1 & 0 & 1
    \end{array} \right) \label{eq:BQCO_BQSO_Ex2}
    \end{align}
 \begin{equation}
BP^{S,N}=BP^{S,O}=\left( \begin{array}{ccc}
         2.55 & 2.5 & 2.6 \\
         1.55 & 1.6 & 1.5 \label{eq:BPSN_BPSO_Ex2}
    \end{array} \right)
\end{equation}
\end{small}

In the non-overbidding case, two matches are established in DP$_2$, between C$_1$ and S$_1$ and C$_2$ and $S_1$ respectively, for 1 unit of gas each, at the price of 2.5. As no participant is overmatched, no bids are dropped; these bids are contracted as well. This results in $TMQ=TCQ=2$, implying a resulting utility value of 2.4 for C$_1$ and C$_2$ each, and a utility value of 0.6 for S$_1$ (thus $TU=5.4$).

What happens in the overbidding case? If we take a look at the figure~\ref{fig:AEU_new_E2_O}, we can see that there is oversupply in DP$_1$ -- for 1 unit of demand, there are overall 3 units of supply. According to the principles of the AEU, the most expensive offers (here, the offer of S$_1$) are disregarded, thus the resulting match will be between $C_1$ and $S_2$, of 1 unit, at the price of 1.55. In DP$_2$, the total quantity of demand and supply offers is the same; two matches are established: one, between C$_1$ and S$_1$, and one between C$_2$ and S$_1$, both of 1 unit, at the price of 2.5. In DP$_3$, the offer of C$_2$ is matched with the bid of $S_2$, regarding 1 unit of gas, at the price of 1.5. The matches resulting from the AEU are summarised in Table \ref{tab:ex_2_matches_by_AEU}.


\begin{table}[t]
    \centering
    \begin{tabular}{c|c|c|c|c|r|c|c|c|c|c|r|c|}
    \cline{2-13}
     & \multicolumn{3}{|c|}{$C_1-S_1$} & \multicolumn{3}{c|}{$C_1-S_2$} & \multicolumn{3}{c|}{$C_2-S_1$} & \multicolumn{3}{c|}{$C_2-S_2$} \\ \hline
 \multicolumn{1}{|c|}{DP} & 1 & 2 & \multicolumn{1}{c|}{3} & 1 & \multicolumn{1}{c|}{2} & \multicolumn{1}{c|}{3} & 1 & 2 & \multicolumn{1}{c|}{3} & 1 & 2 & \multicolumn{1}{c|}{3} \\   \hline
 \multicolumn{1}{|c|}{N}  & 0 & 1 & 0 & 0 & 0 & 0 & 0 & 1 & 0 & 0 & 0 & 0\\
 \multicolumn{1}{|c|}{O}  & 0 & 1 & 0 & 1 & 0 & 0 & 0 & 1 & 0 & 0 & 0 & 1\\\hline
    \end{tabular}
    \caption{Matches by the  (AEU) mechanism in Scenario II for uniform non-overbidding (N) and uniform overbidding (O)}
    \label{tab:ex_2_matches_by_AEU}
\end{table}

We can see that participants C$_1$, C$_2$ and S$_2$ are overmatched. Both C$_1$ and C$_2$ will prefer the cheaper offer of S$_2$ (in DP$_1$ and DP$_3$ respectively) and drop the match of S$_1$ in DP$_2$. S$_2$ will keep the match in DP$_3$ and drop the match in DP$_1$. The only realized (contracted) match is between C$_2$ and S$_2$ in DP$_3$, of 1 unit at the price of 1.5, thus $TMQ=TCQ=1$, implying a resulting utility of 3.3 for C$_2$ and 0.3 for S$_2$, thus $TU=3.6$.

We may note that the universal overbidding strategy reduced the total contracted quantity (TCQ) and the total resulting utility (TU) by 50 and 33.33\% respectively, as compared to the non-overbidding case.

\end{ex}

Looking at the utilities of the two scenarios reveals that uniform overbidding may benefit some players while harming others. In the following subsection, we relax the uniformity assumption and let players choose their bidding strategies.

\subsection{Strategic and equilibrium aspects}
\label{subsec:results_2st_subsec}

While in the previous subsection, we assumed that the bidding strategy of participants is uniform, in this subsection, we consider the case in which participants may individually choose their strategy.

The first observation we can make is the following.
If all other participants follow the non-overbidding strategy, overbidding (O) is at least as beneficial as non-overbidding (N). If only one player is overbidding, he/she may be the only one overmatched, thus the only one who has the potential to choose from the matches according to his/her preferences. Since other players do not drop any bids (since they played `N'), it is certain that the matches chosen according to the overbidding player's preferences will also be realised.

This implies that the uniform non-overbidding strategy is generally not an equilibrium of the implied strategic game (although it may be when nobody gets better options by overbidding). As the simple setup of Scenario I shows in the case of 2 players, universal overbidding may be a unique Nash equilibrium (NE). In this case, if neither player overbids, no matches are established, and the resulting TCQ and utility values will be zero. Furthermore, in this particular case, the strategy `O' is also (weekly) dominant -- if any player switches from `N' to `O', it gets at least as much payoff (resulting utility), as before. An example, where `O' is strictly dominant, may also be easily constructed.

However, as Scenario II (Example~\ref{subsec:Ex2}) shows, switching from `N' to `O' is not necessarily beneficial anymore if some (or all) of the other players are also overbidding. For example, a consumer may obtain better matches (with access to sources at a lower overall cost) through overbidding; however, there is no benefit if these matches are subsequently dropped by the other party.

Table \ref{tab:ex_2_utility_fullStratSpace} summarises the utility values of the players in the case of various strategy profiles, considering Scenario II (the bid and matching Tables are detailed in Appendix A).

\begin{table*}[t]
    \centering
    \begin{tabular}{|c|c|c|c|c|c|c|c|}
\hline
    \multicolumn{1}{|c|}{Strategy} & \multicolumn{1}{c|}{$U^C_1$\rule{0pt}{12pt}} & \multicolumn{1}{c|}{$U^C_2$} &
    \multicolumn{1}{c|}{$U^S_1$} & \multicolumn{1}{c|}{$U^S_2$} &
    \multicolumn{1}{c|}{TMQ} & \multicolumn{1}{c|}{TCQ} & \multicolumn{1}{c|}{TU} \\ \hline
NNNN & 2.4 & 2.4 & 0.6 & 0 & 2 & 2 & 5.4\\
ONNN & 2.4 & 2.4 & 0.6 & 0 & 2 & 2 & 5.4\\
NONN & 2.4 & 3.3 & 0.3 & 0.3 & 3 & 2 & 6.3\\
NNON & 2.4 & 2.4 & 0.6 & 0 & 2 & 2 & 5.4\\
NNNO & 2.4 & 2.4 & 0.6 & 0 & 2 & 2 & 5.4\\
OONN & 2.4 & 3.3 & 0.3 & 0.3 & 3 & 2 & 6.3\\
ONON & 2.4 & 2.4 & 0.6 & 0 & 3 & 2 & 5.4\\
ONNO & 3.25 & 2.4 & 0.3 & 0.25 & 3 & 2 & 6.2\\
NOON & 2.4 & 3.3 & 0.3 & 0.3 & 3 & 2 & 6.3\\
NONO & 2.4 & 3.3 & 0.3 & 0.3 & 3 & 2 & 6.3\\
NNOO & 2.4 & 2.4 & 0.6 & 0 & 2 & 2 & 5.4\\
OOON & 2.4 & 3.3 & 0.3 & 0.3 & 4 & 2 & 6.3\\
OONO & 0 & 3.3 & 0 & 0.3 & 4 & 1 & 3.6\\
ONOO & 3.25 & 2.4 & 0.3 & 0.25 & 3 & 2 & 6.2\\
NOOO & 2.4 & 3.3 & 0.3 & 0.3 & 3 & 2 & 6.3\\
OOOO & 0 & 3.3 & 0 & 0.3 & 4 & 1 & 3.6\\     \hline
\end{tabular}
    \caption{Utility values of individual players, total matched quantity (TMQ), total contracted quantity (TCQ) and total utility in the case of example scenario II for all strategy profiles}
    \label{tab:ex_2_utility_fullStratSpace}
\end{table*}

The `OOOO' strategy profile is not a NE anymore ($C_1$ gets off better by deviating), and overbidding is not always beneficial in this case. This may be seen in the case of other strategy profiles as well, for example in the case of the `NONO' $\rightarrow$ `OONO' transition, where the change of the strategy `N' to `O' by $C_1$ results in a decreased utility ( 2.4 \textit{versus} 0).
The strategy profiles  NONN', `OONN', 'NOON', 'NONO', 'OOON' and 'NOOO' are all NEs, thus this example also illustrates the potential coexistence of multiple equilibria in the implied strategic game.
Furthermore, let us note that while in this particular case all the NEs are efficient in the context of the total traded quantity ($\text{TCQ}=2$, which is the theoretical maximum in this case since $qr^C_1+qr^C_2=2$) and result in the same total utility, this is not necessarily the case (see Scenario III described in Appendix B).

\section{Discussion}

\subsection{The relevance of the model}
To what extent should we be concerned about our findings? What is the applicability of our results to AggregateEU?
While reality presents us with a larger player set and a richer strategy space, the effects captured by our examples are still present; the players we focus on face the described decision problems, potentially resulting in the highlighted inefficiencies.
At the same time, our model has several necessary simplifications, and we now explore the possibility of relaxing them.

In the current model, participants drop unfavourable matches simultaneously. If excess matches are dropped iteratively, one after the other, with plenty of time to inform other players, more previously less preferred deals are also kept. This would imply more realised matches compared to the results provided by our model.

In reality, the exact time when a deal is dropped may get blurred by, for example, additional rounds of negotiations before opting out, not to mention challenges to transfer this information into common knowledge.
Taking this iterative aspect of bargaining into account would add much to the complexity of the model and make calculations excessively challenging.

Likewise, allowing for a richer set of strategies for overbidding beyond the 100\% studied here would make the model more realistic, but also substantially more complex.



\subsection{Characterising the potential effects of overbidding and inefficient scenarios}

Based on the two scenarios presented, we may make the following observations.
\begin{itemize}
    \item Overbidding may happen only if the participant in question has enough total access capacity to DPs. Thus, if the total access of each participant to DPs is less than or equal to their required or available quantity.
    \item If the preference lists of all participants over the DPs are the same, overbidding will have no effect. Since total supply and demand will already meet at the most preferred DPs, further matches at less preferred DPs will be dropped by both consumers and suppliers.
    \item As Scenario I described in Example \ref{subsec:Ex1} shows, overbidding may be beneficial in setups where the preference lists of consumers and suppliers over DPs are opposite. In this case, placing bids on less preferred DPs may help establish matches between supply and demand, allowing for trading that would not occur otherwise.
    \item As illustrated by Scenario II described in Example \ref{subsec:Ex2}, the necessary condition for inefficiency in the overbidding case is that at least 1 consumer and at least 1 producer have to be overmatched in the process; otherwise, the asymmetric drop of matches, which is the essence of inefficiency, can not happen.
\end{itemize}

While a comprehensive analysis fully characterising the gas markets under the AggregateEU mechanism is beyond the scope of this paper, in the following, we highlight a common conflict of interest in two-sided platforms. Indeed, while AggregateEU coordinates the entire mechanism, trade takes place at the delivery points that, as a result, act as platforms.

Seminal papers on two-sided platforms show that the optimal business strategies on platforms can differ fundamentally from those in classic
markets \citep{Armstrong2006, CaillaudJullien2003, RochetTirole2003, RochetTirole2006}. More recent contributions to this literature
 explore the role of bounded rationality of some agents  \citep{JohnenSomogyi2024} and the degree of competition
between sellers \citep{BelleflammePeitz2019, KarlePeitzReisinger2020}.

In the case of the AggregateEU mechanism, some of the miscoordination is due to the opposing preferences of buyers and sellers in a multihoming setup \citep{BakosHalaburda2020}. For illustration, consider the following example: A shop offers cash or card payments, which are the two platforms for the transaction. Cash payments are cumbersome for the buyer, while card payments have higher costs for the shop. Under the current mechanism both options are dropped, as the shop prefers cash --- and drops the card option --- the customer prefers card payment --- and drops cash as the means of payment.

This type of miscoordination is a common problem for platforms as well, not least because of the pricing strategies employed by the platforms themselves for both sides. If, for example, credit card users get rebates, customers stop carrying cash, and the shops must accept them to complete the transaction over a less favourable platform. Here, consumers and/or suppliers must take the other side's preferences into account when dropping matches.

In the following subsection, we present some modified algorithms to improve efficiency.

\subsection{Alternative mechanisms}
\label{subsec_alternatives}

While overbidding can be an efficient tool to link supply and demand, we find that it may also backfire due to the non-binding nature of the AEU, which allows matches to drop. If the policymakers' aim is to get rid of these potential inefficiencies, or attenuate their effect, the following options may be considered.

\paragraph{Convex bid sets} Even if the nonbinding nature of the bids is essentially required, we may improve efficiency without overbidding and the related outcome uncertainty.
\textit{Convex bid sets} \citep{csercsik2022convex}, allow participants to bid at several different DPs an overall quantity greater than $qr^C$ or $qa^S$ while also avoiding overmatching. In this case, the clearing algorithm must take into account that the individual bids are parts of a convex bid set and determine their acceptance/rejection accordingly, i.e., by not allowing the overmatching of any participant. This may be straightforward in some cases (when some of a supplier's bids are present in a DP where there is oversupply, or vice versa for consumers), but in general, the choice of matched and non-matched (cancelled) bids is highly non-trivial. For example, cancelling a supply bid at a DP (e.g., at a less preferred DP of a supplier) will decrease the total supply at the corresponding DP, thus also limiting the demand bids that may be matched there. Even if the total matched quantity remains unaffected, choosing the matches potentially affects the resulting utility of participants by determining which consumers gain access to cheaper sources (even if the pro-rata principle is applied at each DP), and how welfare is distributed among participants. Analysing axiomatic approaches in such multi-platform trading markets, which incorporate convex bid sets, may be a promising future research topic.

\paragraph{Delivery constraints} The current AggregateEU mechanism does not account for the capacity constraints of LNG terminals (DPs) in the matching process. In the absence of such a constraint, a match using one or more DPs beyond their physical capacities is theoretically possible. A clearing mechanism with complex bid sets and DP-related constraints could avoid this.


\paragraph{Binding mechanisms} If partially or fully binding mechanisms are options, there are multiple alternatives.

(i) The first possibility is to complement the non-binding round of the AggregateEU with a binding second round, in which participants may update their previous offers subject to limitations (for example, by not more than 50\% of the first round offers), but the results of this second round are already binding. This concept may be, of course, generalised for multiple rounds as well, which would imply a more cumbersome bidding process, but could potentially enhance the convergence to an efficient solution. The concept of combinatorial clock auctions \cite{ausubel2014market} may be relevant in the context of such a mechanism.

(ii) Alternatively, the first round could already be binding.
A binding framework makes overbidding impossible (if penalties for non-delivery are high enough).
Such a framework should include more complex offer types incorporating alternatives for consumers and suppliers in the context of DPs, such as the convex bid sets discussed above. A single-round binding mechanism should also include ways to avoid matches that may turn out too expensive for consumers in the contracting phase, for example, by allowing buyers to set a reservation price or negotiate deals.

\paragraph{No overbidding for suppliers} Overbidding under the AggregateEU mechanism can lead to inefficiencies in an unfortunate combination with miscoordinated drops of matches. Removing the coordination problem by disallowing suppliers to overbid removes the uncertainty at the heart of the problem. When suppliers cannot overbid, they have no interest in dropping the matches found by the mechanism; therefore, buyers can count on the realisation of all the matches. As buyers keep their matches up to their demand, the total contracted quantity TCQ would be maximal, and even the costs under the realised contracts would be near optimal due to the buyers' preferences. In equilibrium, buyers would select suppliers with the lowest total prices, while a coordination failure across DPs may persist, with some DPs only served by high-cost suppliers.

\paragraph{Delayed acceptance of offers} The Gale-Shapley or \textit{delayed acceptance algorithm} \citep{gale1962college} has been introduced for the marriage market but it is fairly easy to generalise it for the current problem, assuming every participant may define a strict ordering on its matches, and no ties appear (like in the proposed model implied by equal utility values).
\begin{description}
    \item[Step 0] The Aggregate EU mechanism establishes a list of matches specifying the parties, the quantities and, in the case of the sellers, the prices.
    \item[Step 1] Consumers each rank the matches and signal their intent to contract on the most favoured ones up to their capacities, taking both the capacities to the delivery points and the overall required quantity into account. Such offers will typically include a last match where the offer is only about a fraction of the matched quantity.
    \item[Step 2] Suppliers rank the offers and select the most favoured ones up to their capacities, taking both the capacities to the delivery points and the overall supply capacity into account. These offers are kept. Again, the kept offers will typically include a last one where only a fraction of the offered quantity is kept, and the rest gets rejected. Rejected matches (or parts thereof) are dropped for good.
    \item[Step $n$] Consumers whose offers got rejected each rank the unused matches and signal their intent to contract on the most favoured ones up to their remaining capacities (that is, the total capacity rejected by sellers), taking capacity constraints into account. If no unused matches are available, the algorithm stops.
    \item[Step $n+1$] Suppliers compare these new offers and the ones kept in the previous round and select the most favoured ones up to their capacities, taking capacity constraints into account. These offers are kept, while the rest are rejected. If no offers get rejected, the algorithm stops; otherwise, it continues at Step $n$.
\end{description}
The stable allocation problem was originally introduced in \citep{BaiouBalinski2002} and generalised in \citep{BiroFleiner2010}, with more efficient algorithms provided by \cite{DeanMunshi2010}.

Due to the well-established properties of the algorithm, the resulting contracts are stable. Since this version is the consumer-proposing algorithm, the contracts are optimal for the consumers. There is a natural counterpart where the producers propose, with the final contracts being optimal for them.

This approach requires suppliers to rank matches. While suppliers may easily rank matches corresponding to various DPs based on the respective transfer costs, it is not trivial how a supplier would rank multiple matches within the same DP, which could limit the applicability of such a method.


\section{Conclusions}
In this paper, we study the overbidding phenomenon in the framework of the AggregateEU mechanism. Using a simple quantitative model, we show that overbidding, which is possible in the current design, has an ambiguous effect on the outcomes. On the one hand, it can facilitate trade and increase market efficiency by helping to connect otherwise unmatched supply and demand. On the other hand, an unfortunate combination of overmatching and the uncoordinated drop of excess matches (as illustrated by Scenario II) can result in efficiency loss, where the total contracted quantity ($TCQ$) falls below the uniform non-overbidding case despite a higher total quantity of matches established by the platform.

In this analysis, we \textit{assume} that market participants are overbidding. Seeing that this is not always in their interest, we study the AggregateEU mechanism with overbidding as a strategic choice. We analyse the strategy space of the implied noncooperative game, where participants can choose to bid truthfully or overbid, and study two parameterisations (Scenarios I and II). The results show that, while it is possible that overbidding is a dominant strategy resulting in an efficient Nash equilibrium, multiple equilibria with significantly different efficiency may also arise.

Finally, in Subsection \ref{subsec_alternatives}, we discuss several alternatives. While even the current non-binding type framework can be improved with the introduction of convex bid sets or restrictions on overbidding, partially or fully binding alternatives may also lead to efficiency gains. Such options should be considered with great caution, as our goal is not to reinvent traditional markets but to create additional trading platforms.
The optimal solution would be an innovative, albeit complex, trading framework that incorporates and coordinates multiple bid types, including conventional bid types of natural gas exchanges and novel convex bids over multiple DPs as well, and provides a unified clearing solution ensuring flexibility, efficiency, and fairness.

\section{Acknowledgements}
The authors thank discussions with Péter Biró, Róbert Somogyi, and Szilvia Pápai.
This work has been supported by the Hungarian Academy of Sciences under its Momentum Programme (LP2021-2) and by Grant K146320 of the Hungarian National Research, Development, and Innovation Office.

\bibliography{bibs}

\appendix

\newpage
\section*{Appendix A: Supporting calculations for Scenario II}
Bids and matchings of Scenario II are detailed in Example \ref{subsec:Ex2} for the complete strategy space.

\setlength\tabcolsep{3pt} 
\begin{table*}[t]
    \centering
    \begin{scriptsize}
    \begin{tabular}{c|c|c|c|c|c|c|c|c|c|c|c|c|c|c|c|c|c|c|}
    \cline{2-19}
    & \multicolumn{12}{|c|}{Quantity} & \multicolumn{6}{|c|}{Price} \\ \cline{2-19}
    & \multicolumn{3}{|c|}{$C_1$} & \multicolumn{3}{|c|}{$C_2$} & \multicolumn{3}{|c|}{$S_1$} & \multicolumn{3}{|c|}{$S_2$} & \multicolumn{3}{|c|}{$S_1$} & \multicolumn{3}{|c|}{$S_2$} \\ \hline
    \multicolumn{1}{|c|}{\backslashbox{SP}{DP}} & 1 & 2 & 3 & 1 & 2 & 3 & 1 & 2 & 3 & 1 & 2 & 3 & 1 & 2 & 3 & 1 & 2 & 3 \\ \hline
\multicolumn{1}{|c|}{NNNN} & 0 & 1 & 0 & 0 & 1 & 0 & 0 & 2 & 0 & 0 & 0 & 1 & 0 & 2.5 & 0 & 0 & 0 & 1.5\\
\multicolumn{1}{|c|}{ONNN} & 1 & 1 & 0 & 0 & 1 & 0 & 0 & 2 & 0 & 0 & 0 & 1 & 0 & 2.5 & 0 & 0 & 0 & 1.5\\
\multicolumn{1}{|c|}{NONN} & 0 & 1 & 0 & 0 & 1 & 1 & 0 & 2 & 0 & 0 & 0 & 1 & 0 & 2.5 & 0 & 0 & 0 & 1.5\\
\multicolumn{1}{|c|}{NNON} & 0 & 1 & 0 & 0 & 1 & 0 & 2 & 2 & 0 & 0 & 0 & 1 & 2.55 & 2.5 & 0 & 0 & 0 & 1.5\\
\multicolumn{1}{|c|}{NNNO} & 0 & 1 & 0 & 0 & 1 & 0 & 0 & 2 & 0 & 1 & 0 & 1 & 0 & 2.5 & 0 & 1.55 & 0 & 1.5\\
\multicolumn{1}{|c|}{OONN} & 1 & 1 & 0 & 0 & 1 & 1 & 0 & 2 & 0 & 0 & 0 & 1 & 0 & 2.5 & 0 & 0 & 0 & 1.5\\
\multicolumn{1}{|c|}{ONON} & 1 & 1 & 0 & 0 & 1 & 0 & 2 & 2 & 0 & 0 & 0 & 1 & 2.55 & 2.5 & 0 & 0 & 0 & 1.5\\
\multicolumn{1}{|c|}{ONNO} & 1 & 1 & 0 & 0 & 1 & 0 & 0 & 2 & 0 & 1 & 0 & 1 & 0 & 2.5 & 0 & 1.55 & 0 & 1.5\\
\multicolumn{1}{|c|}{NOON} & 0 & 1 & 0 & 0 & 1 & 1 & 2 & 2 & 0 & 0 & 0 & 1 & 2.55 & 2.5 & 0 & 0 & 0 & 1.5\\
\multicolumn{1}{|c|}{NONO} & 0 & 1 & 0 & 0 & 1 & 1 & 0 & 2 & 0 & 1 & 0 & 1 & 0 & 2.5 & 0 & 1.55 & 0 & 1.5\\
\multicolumn{1}{|c|}{NNOO} & 0 & 1 & 0 & 0 & 1 & 0 & 2 & 2 & 0 & 1 & 0 & 1 & 2.55 & 2.5 & 0 & 1.55 & 0 & 1.5\\
\multicolumn{1}{|c|}{OOON} & 1 & 1 & 0 & 0 & 1 & 1 & 2 & 2 & 0 & 0 & 0 & 1 & 2.55 & 2.5 & 0 & 0 & 0 & 1.5\\
\multicolumn{1}{|c|}{OONO} & 1 & 1 & 0 & 0 & 1 & 1 & 0 & 2 & 0 & 1 & 0 & 1 & 0 & 2.5 & 0 & 1.55 & 0 & 1.5\\
\multicolumn{1}{|c|}{ONOO} & 1 & 1 & 0 & 0 & 1 & 0 & 2 & 2 & 0 & 1 & 0 & 1 & 2.55 & 2.5 & 0 & 1.55 & 0 & 1.5\\
\multicolumn{1}{|c|}{NOOO} & 0 & 1 & 0 & 0 & 1 & 1 & 2 & 2 & 0 & 1 & 0 & 1 & 2.55 & 2.5 & 0 & 1.55 & 0 & 1.5\\
\multicolumn{1}{|c|}{OOOO} & 1 & 1 & 0 & 0 & 1 & 1 & 2 & 2 & 0 & 1 & 0 & 1 & 2.55 & 2.5 & 0 & 1.55 & 0 & 1.5\\     \hline
    \end{tabular}
    \end{scriptsize}
    \caption{Bids of Scenario II in the case of various strategy profiles (SPs).}
    \label{tab:Bids_Ex_2_fullStratSpace}
\end{table*}
\setlength\tabcolsep{6pt} 

\setlength\tabcolsep{4.6pt} 
\begin{table*}[t]
    \centering
    \begin{tabular}{c|c|r|c|c|r|c|c|r|c|c|r|c|}
    \cline{2-13}
     & \multicolumn{3}{|c|}{$C_1-S_1$} & \multicolumn{3}{c|}{$C_1-S_2$} & \multicolumn{3}{c|}{$C_2-S_1$} & \multicolumn{3}{c|}{$C_2-S_2$} \\ \hline
    \multicolumn{1}{|c|}{{\backslashbox{SP}{DP}}} & \multicolumn{1}{c}{1} & \multicolumn{1}{c}{2} & \multicolumn{1}{c|}{3} &
                               \multicolumn{1}{c}{1} & \multicolumn{1}{c}{2} & \multicolumn{1}{c|}{3} &
                               \multicolumn{1}{c}{1} & \multicolumn{1}{c}{2} & \multicolumn{1}{c|}{3} &
                               \multicolumn{1}{c}{1} & \multicolumn{1}{c}{2} &
                               \multicolumn{1}{c|}{3}\\\hline
\multicolumn{1}{|c|}{NNNN} & 0 & 1 & 0 & 0 & 0 & 0 & 0 & 1 & 0 & 0 & 0 & 0\\
\multicolumn{1}{|c|}{ONNN} & 0 & 1 & 0 & 0 & 0 & 0 & 0 & 1 & 0 & 0 & 0 & 0\\
\multicolumn{1}{|c|}{NONN} & 0 & 1 & 0 & 0 & 0 & 0 & 0 & 1 & 0 & 0 & 0 & 1\\
\multicolumn{1}{|c|}{NNON} & 0 & 1 & 0 & 0 & 0 & 0 & 0 & 1 & 0 & 0 & 0 & 0\\
\multicolumn{1}{|c|}{NNNO} & 0 & 1 & 0 & 0 & 0 & 0 & 0 & 1 & 0 & 0 & 0 & 0\\
\multicolumn{1}{|c|}{OONN} & 0 & 1 & 0 & 0 & 0 & 0 & 0 & 1 & 0 & 0 & 0 & 1\\
\multicolumn{1}{|c|}{ONON} & 1 & 1 & 0 & 0 & 0 & 0 & 0 & 1 & 0 & 0 & 0 & 0\\
\multicolumn{1}{|c|}{ONNO} & 0 & 1 & 0 & 1 & 0 & 0 & 0 & 1 & 0 & 0 & 0 & 0\\
\multicolumn{1}{|c|}{NOON} & 0 & 1 & 0 & 0 & 0 & 0 & 0 & 1 & 0 & 0 & 0 & 1\\
\multicolumn{1}{|c|}{NONO} & 0 & 1 & 0 & 0 & 0 & 0 & 0 & 1 & 0 & 0 & 0 & 1\\
\multicolumn{1}{|c|}{NNOO} & 0 & 1 & 0 & 0 & 0 & 0 & 0 & 1 & 0 & 0 & 0 & 0\\
\multicolumn{1}{|c|}{OOON} & 1 & 1 & 0 & 0 & 0 & 0 & 0 & 1 & 0 & 0 & 0 & 1\\
\multicolumn{1}{|c|}{OONO} & 0 & 1 & 0 & 1 & 0 & 0 & 0 & 1 & 0 & 0 & 0 & 1\\
\multicolumn{1}{|c|}{ONOO} & 0 & 1 & 0 & 1 & 0 & 0 & 0 & 1 & 0 & 0 & 0 & 0\\
\multicolumn{1}{|c|}{NOOO} & 0 & 1 & 0 & 0 & 0 & 0 & 0 & 1 & 0 & 0 & 0 & 1\\
\multicolumn{1}{|c|}{OOOO} & 0 & 1 & 0 & 1 & 0 & 0 & 0 & 1 & 0 & 0 & 0 & 1\\
 \hline
    \end{tabular}
    \caption{matches by AEU in the case of Scenario II in the case of various strategy profiles (SPs).}
    \label{tab:Results_ex2_1_fullStratSpace}
\end{table*}
\setlength\tabcolsep{6pt} 

\begin{table*}[t]
    \centering
    \begin{tabular}{c|r|r|c|c|c|c|r|r|c|c|r|r|}
    \cline{2-13}
     & \multicolumn{3}{|c|}{$C_1-S_1$} & \multicolumn{3}{c|}{$C_1-S_2$} & \multicolumn{3}{c|}{$C_2-S_1$} & \multicolumn{3}{c|}{$C_2-S_2$} \\ \hline
    \multicolumn{1}{|c|}{{\backslashbox{SP}{DP}}} & \multicolumn{1}{c}{1} & \multicolumn{1}{c}{2} & \multicolumn{1}{c|}{3} &
                               \multicolumn{1}{c}{1} & \multicolumn{1}{c}{2} & \multicolumn{1}{c|}{3} &
                               \multicolumn{1}{c}{1} & \multicolumn{1}{c}{2} & \multicolumn{1}{c|}{3} &
                               \multicolumn{1}{c}{1} & \multicolumn{1}{c}{2} &
                               \multicolumn{1}{c|}{3}\\\hline
\multicolumn{1}{|c|}{NNNN} & 0 & 1 & 0 & 0 & 0 & 0 & 0 & 1 & 0 & 0 & 0 & 0\\
\multicolumn{1}{|c|}{ONNN} & 0 & 1 & 0 & 0 & 0 & 0 & 0 & 1 & 0 & 0 & 0 & 0\\
\multicolumn{1}{|c|}{NONN} & 0 & 1 & 0 & 0 & 0 & 0 & 0 & 0 & 0 & 0 & 0 & 1\\
\multicolumn{1}{|c|}{NNON} & 0 & 1 & 0 & 0 & 0 & 0 & 0 & 1 & 0 & 0 & 0 & 0\\
\multicolumn{1}{|c|}{NNNO} & 0 & 1 & 0 & 0 & 0 & 0 & 0 & 1 & 0 & 0 & 0 & 0\\
\multicolumn{1}{|c|}{OONN} & 0 & 1 & 0 & 0 & 0 & 0 & 0 & 0 & 0 & 0 & 0 & 1\\
\multicolumn{1}{|c|}{ONON} & 0 & 1 & 0 & 0 & 0 & 0 & 0 & 1 & 0 & 0 & 0 & 0\\
\multicolumn{1}{|c|}{ONNO} & 0 & 0 & 0 & 1 & 0 & 0 & 0 & 1 & 0 & 0 & 0 & 0\\
\multicolumn{1}{|c|}{NOON} & 0 & 1 & 0 & 0 & 0 & 0 & 0 & 0 & 0 & 0 & 0 & 1\\
\multicolumn{1}{|c|}{NONO} & 0 & 1 & 0 & 0 & 0 & 0 & 0 & 0 & 0 & 0 & 0 & 1\\
\multicolumn{1}{|c|}{NNOO} & 0 & 1 & 0 & 0 & 0 & 0 & 0 & 1 & 0 & 0 & 0 & 0\\
\multicolumn{1}{|c|}{OOON} & 0 & 1 & 0 & 0 & 0 & 0 & 0 & 0 & 0 & 0 & 0 & 1\\
\multicolumn{1}{|c|}{OONO} & 0 & 0 & 0 & 0 & 0 & 0 & 0 & 0 & 0 & 0 & 0 & 1\\
\multicolumn{1}{|c|}{ONOO} & 0 & 0 & 0 & 1 & 0 & 0 & 0 & 1 & 0 & 0 & 0 & 0\\
\multicolumn{1}{|c|}{NOOO} & 0 & 1 & 0 & 0 & 0 & 0 & 0 & 0 & 0 & 0 & 0 & 1\\
\multicolumn{1}{|c|}{OOOO} & 0 & 0 & 0 & 0 & 0 & 0 & 0 & 0 & 0 & 0 & 0 & 1\\
 \hline
    \end{tabular}
    \caption{Realised (contracted) matches in the case of Scenario II in the case of various strategy profiles (SPs).}
    \label{tab:Results_ex2_2_fullStratSpace}
\end{table*}


\newpage
\section*{Appendix B: Multiple Nash equilibria with different efficiency (Scenario III)}
Parameters are the same as in the case of Scenario II, described by eqs. (\ref{eq_par_cons_2}) and (\ref{eq_par_sup_2}), except for $qr^C_1$, which is increased from 1 to 1.5. As it may be seen in Table \ref{tab:Results_ex3_1_fullStratSpace},
, 'OONN', 'OOON', 'NONO' and 'NOOO' are all NEs, but the former two imply $\text{TCQ}=2.5$ and $TU=7.6$  while the latter two result in $\text{TCQ}=2$ and $TU=6.3$.

\begin{align}
     qr^C_1&= 1.5 \quad &u^C_1&=5
     \nonumber \\\overline{q}^C_1&= [1~~ 1~~1] \quad &ct^C_1&=[0.2~~ 0.1~~0.3] \nonumber \\
     qr^C_2&= 1 \quad &u^C_2&=5
     \nonumber \\\overline{q}^C_2&= [1~~ 1~~1] \quad &ct^C_2&=[0.3~~ 0.1~~0.2]
     \label{eq_par_cons_3} \\
    qa^S_1&= 2 \quad &cp^S_1&=2 \nonumber \\ \overline{q}^S_1&= [2~~ 2~~2] \quad &ct^S_1&=[0.3~~ 0.2~~ 0.4]  \nonumber \\
    qa^S_2&= 1 \quad &cp^S_2&=1 \nonumber \\ \overline{q}^S_2&= [1~~ 1~~1] \quad &ct^S_2&=[0.3~~ 0.4~~ 0.2]\label{eq_par_sup_3}
\end{align}

\setlength\tabcolsep{3pt} 
\begin{table*}[t]
    \centering
    \begin{scriptsize}
    \begin{tabular}{c|c|c|c|c|c|c|c|c|c|c|c|c|c|c|c|c|c|c|}
    \cline{2-19}
    & \multicolumn{12}{|c|}{Quantity} & \multicolumn{6}{|c|}{Price} \\ \cline{2-19}
    & \multicolumn{3}{|c|}{$C_1$} & \multicolumn{3}{|c|}{$C_2$} & \multicolumn{3}{|c|}{$S_1$} & \multicolumn{3}{|c|}{$S_2$} & \multicolumn{3}{|c|}{$S_1$} & \multicolumn{3}{|c|}{$S_2$} \\ \hline
    \multicolumn{1}{|c|}{\backslashbox{SP}{DP}} & 1 & 2 & 3 & 1 & 2 & 3 & 1 & 2 & 3 & 1 & 2 & 3 & 1 & 2 & 3 & 1 & 2 & 3 \\ \hline
\multicolumn{1}{|c|}{NNNN} & 0.5 & 1 & 0 & 0 & 1 & 0 & 0 & 2 & 0 & 0 & 0 & 1 & 0 & 2.5 & 0 & 0 & 0 & 1.5\\
\multicolumn{1}{|c|}{ONNN} & 1 & 1 & 1 & 0 & 1 & 0 & 0 & 2 & 0 & 0 & 0 & 1 & 0 & 2.5 & 0 & 0 & 0 & 1.5\\
\multicolumn{1}{|c|}{NONN} & 0.5 & 1 & 0 & 0 & 1 & 1 & 0 & 2 & 0 & 0 & 0 & 1 & 0 & 2.5 & 0 & 0 & 0 & 1.5\\
\multicolumn{1}{|c|}{NNON} & 0.5 & 1 & 0 & 0 & 1 & 0 & 2 & 2 & 0 & 0 & 0 & 1 & 2.55 & 2.5 & 0 & 0 & 0 & 1.5\\
\multicolumn{1}{|c|}{NNNO} & 0.5 & 1 & 0 & 0 & 1 & 0 & 0 & 2 & 0 & 1 & 0 & 1 & 0 & 2.5 & 0 & 1.55 & 0 & 1.5\\
\multicolumn{1}{|c|}{OONN} & 1 & 1 & 1 & 0 & 1 & 1 & 0 & 2 & 0 & 0 & 0 & 1 & 0 & 2.5 & 0 & 0 & 0 & 1.5\\
\multicolumn{1}{|c|}{ONON} & 1 & 1 & 1 & 0 & 1 & 0 & 2 & 2 & 0 & 0 & 0 & 1 & 2.55 & 2.5 & 0 & 0 & 0 & 1.5\\
\multicolumn{1}{|c|}{ONNO} & 1 & 1 & 1 & 0 & 1 & 0 & 0 & 2 & 0 & 1 & 0 & 1 & 0 & 2.5 & 0 & 1.55 & 0 & 1.5\\
\multicolumn{1}{|c|}{NOON} & 0.5 & 1 & 0 & 0 & 1 & 1 & 2 & 2 & 0 & 0 & 0 & 1 & 2.55 & 2.5 & 0 & 0 & 0 & 1.5\\
\multicolumn{1}{|c|}{NONO} & 0.5 & 1 & 0 & 0 & 1 & 1 & 0 & 2 & 0 & 1 & 0 & 1 & 0 & 2.5 & 0 & 1.55 & 0 & 1.5\\
\multicolumn{1}{|c|}{NNOO} & 0.5 & 1 & 0 & 0 & 1 & 0 & 2 & 2 & 0 & 1 & 0 & 1 & 2.55 & 2.5 & 0 & 1.55 & 0 & 1.5\\
\multicolumn{1}{|c|}{OOON} & 1 & 1 & 1 & 0 & 1 & 1 & 2 & 2 & 0 & 0 & 0 & 1 & 2.55 & 2.5 & 0 & 0 & 0 & 1.5\\
\multicolumn{1}{|c|}{OONO} & 1 & 1 & 1 & 0 & 1 & 1 & 0 & 2 & 0 & 1 & 0 & 1 & 0 & 2.5 & 0 & 1.55 & 0 & 1.5\\
\multicolumn{1}{|c|}{ONOO} & 1 & 1 & 1 & 0 & 1 & 0 & 2 & 2 & 0 & 1 & 0 & 1 & 2.55 & 2.5 & 0 & 1.55 & 0 & 1.5\\
\multicolumn{1}{|c|}{NOOO} & 0.5 & 1 & 0 & 0 & 1 & 1 & 2 & 2 & 0 & 1 & 0 & 1 & 2.55 & 2.5 & 0 & 1.55 & 0 & 1.5\\
\multicolumn{1}{|c|}{OOOO} & 1 & 1 & 1 & 0 & 1 & 1 & 2 & 2 & 0 & 1 & 0 & 1 & 2.55 & 2.5 & 0 & 1.55 & 0 & 1.5\\     \hline
    \end{tabular}
    \end{scriptsize}
    \caption{Bids of Scenario III in the case of various strategy profiles (SPs).}
    \label{tab:Bids_Ex_3_fullStratSpace}
\end{table*}
\setlength\tabcolsep{6pt} 

\setlength\tabcolsep{4.6pt} 
\begin{table*}[t]
    \centering
    \begin{tabular}{c|c|r|c|c|r|c|c|r|c|c|r|c|}
    \cline{2-13}
     & \multicolumn{3}{|c|}{$C_1-S_1$} & \multicolumn{3}{c|}{$C_1-S_2$} & \multicolumn{3}{c|}{$C_2-S_1$} & \multicolumn{3}{c|}{$C_2-S_2$} \\ \hline
    \multicolumn{1}{|c|}{{\backslashbox{SP}{DP}}} & \multicolumn{1}{c}{1} & \multicolumn{1}{c}{2} & \multicolumn{1}{c|}{3} &
    \multicolumn{1}{c}{1} & \multicolumn{1}{c}{2} & \multicolumn{1}{c|}{3} &
    \multicolumn{1}{c}{1} & \multicolumn{1}{c}{2} & \multicolumn{1}{c|}{3} &
    \multicolumn{1}{c}{1} & \multicolumn{1}{c}{2} &
    \multicolumn{1}{c|}{3}\\\hline
\multicolumn{1}{|c|}{NNNN} & 0 & 1 & 0 & 0 & 0 & 0 & 0 & 1 & 0 & 0 & 0 & 0\\
\multicolumn{1}{|c|}{ONNN} & 0 & 1 & 0 & 0 & 0 & 1 & 0 & 1 & 0 & 0 & 0 & 0\\
\multicolumn{1}{|c|}{NONN} & 0 & 1 & 0 & 0 & 0 & 0 & 0 & 1 & 0 & 0 & 0 & 1\\
\multicolumn{1}{|c|}{NNON} & 0.5 & 1 & 0 & 0 & 0 & 0 & 0 & 1 & 0 & 0 & 0 & 0\\
\multicolumn{1}{|c|}{NNNO} & 0 & 1 & 0 & 0.5 & 0 & 0 & 0 & 1 & 0 & 0 & 0 & 0\\
\multicolumn{1}{|c|}{OONN} & 0 & 1 & 0 & 0 & 0 & 0.5 & 0 & 1 & 0 & 0 & 0 & 0.5\\
\multicolumn{1}{|c|}{ONON} & 1 & 1 & 0 & 0 & 0 & 1 & 0 & 1 & 0 & 0 & 0 & 0\\
\multicolumn{1}{|c|}{ONNO} & 0 & 1 & 0 & 1 & 0 & 1 & 0 & 1 & 0 & 0 & 0 & 0\\
\multicolumn{1}{|c|}{NOON} & 0.5 & 1 & 0 & 0 & 0 & 0 & 0 & 1 & 0 & 0 & 0 & 1\\
\multicolumn{1}{|c|}{NONO} & 0 & 1 & 0 & 0.5 & 0 & 0 & 0 & 1 & 0 & 0 & 0 & 1\\
\multicolumn{1}{|c|}{NNOO} & 0 & 1 & 0 & 0.5 & 0 & 0 & 0 & 1 & 0 & 0 & 0 & 0\\
\multicolumn{1}{|c|}{OOON} & 1 & 1 & 0 & 0 & 0 & 0.5 & 0 & 1 & 0 & 0 & 0 & 0.5\\
\multicolumn{1}{|c|}{OONO} & 0 & 1 & 0 & 1 & 0 & 0.5 & 0 & 1 & 0 & 0 & 0 & 0.5\\
\multicolumn{1}{|c|}{ONOO} & 0 & 1 & 0 & 1 & 0 & 1 & 0 & 1 & 0 & 0 & 0 & 0\\
\multicolumn{1}{|c|}{NOOO} & 0 & 1 & 0 & 0.5 & 0 & 0 & 0 & 1 & 0 & 0 & 0 & 1\\
\multicolumn{1}{|c|}{OOOO} & 0 & 1 & 0 & 1 & 0 & 0.5 & 0 & 1 & 0 & 0 & 0 & 0.5\\
 \hline
    \end{tabular}
    \caption{matches by AEU in the case of Scenario III in the case of various strategy profiles (SPs).}
    \label{tab:Results_ex3_1_fullStratSpace}
\end{table*}
\setlength\tabcolsep{6pt} 

\begin{table*}[t]
    \centering
    \begin{tabular}{c|r|r|c|c|c|c|r|r|c|c|r|r|}
    \cline{2-13}
     & \multicolumn{3}{|c|}{$C_1-S_1$} & \multicolumn{3}{c|}{$C_1-S_2$} & \multicolumn{3}{c|}{$C_2-S_1$} & \multicolumn{3}{c|}{$C_2-S_2$} \\ \hline
    \multicolumn{1}{|c|}{{\backslashbox{SP}{DP}}} & \multicolumn{1}{c}{1} & \multicolumn{1}{c}{2} & \multicolumn{1}{c|}{3} &
                               \multicolumn{1}{c}{1} & \multicolumn{1}{c}{2} & \multicolumn{1}{c|}{3} &
                               \multicolumn{1}{c}{1} & \multicolumn{1}{c}{2} & \multicolumn{1}{c|}{3} &
                               \multicolumn{1}{c}{1} & \multicolumn{1}{c}{2} &
                               \multicolumn{1}{c|}{3}\\\hline
\multicolumn{1}{|c|}{NNNN} & 0 & 1 & 0 & 0 & 0 & 0 & 0 & 1 & 0 & 0 & 0 & 0\\
\multicolumn{1}{|c|}{ONNN} & 0 & 0.5 & 0 & 0 & 0 & 1 & 0 & 1 & 0 & 0 & 0 & 0\\
\multicolumn{1}{|c|}{NONN} & 0 & 1 & 0 & 0 & 0 & 0 & 0 & 0 & 0 & 0 & 0 & 1\\
\multicolumn{1}{|c|}{NNON} & 0 & 1 & 0 & 0 & 0 & 0 & 0 & 1 & 0 & 0 & 0 & 0\\
\multicolumn{1}{|c|}{NNNO} & 0 & 1 & 0 & 0.5 & 0 & 0 & 0 & 1 & 0 & 0 & 0 & 0\\
\multicolumn{1}{|c|}{OONN} & 0 & 1 & 0 & 0 & 0 & 0.5 & 0 & 0.5 & 0 & 0 & 0 & 0.5\\
\multicolumn{1}{|c|}{ONON} & 0 & 0.5 & 0 & 0 & 0 & 1 & 0 & 1 & 0 & 0 & 0 & 0\\
\multicolumn{1}{|c|}{ONNO} & 0 & 0 & 0 & 0 & 0 & 0.5 & 0 & 1 & 0 & 0 & 0 & 0\\
\multicolumn{1}{|c|}{NOON} & 0 & 1 & 0 & 0 & 0 & 0 & 0 & 0 & 0 & 0 & 0 & 1\\
\multicolumn{1}{|c|}{NONO} & 0 & 1 & 0 & 0 & 0 & 0 & 0 & 0 & 0 & 0 & 0 & 1\\
\multicolumn{1}{|c|}{NNOO} & 0 & 1 & 0 & 0.5 & 0 & 0 & 0 & 1 & 0 & 0 & 0 & 0\\
\multicolumn{1}{|c|}{OOON} & 0 & 1 & 0 & 0 & 0 & 0.5 & 0 & 0.5 & 0 & 0 & 0 & 0.5\\
\multicolumn{1}{|c|}{OONO} & 0 & 0 & 0 & 0 & 0 & 0.5 & 0 & 0.5 & 0 & 0 & 0 & 0.5\\
\multicolumn{1}{|c|}{ONOO} & 0 & 0 & 0 & 0 & 0 & 0.5 & 0 & 1 & 0 & 0 & 0 & 0\\
\multicolumn{1}{|c|}{NOOO} & 0 & 1 & 0 & 0 & 0 & 0 & 0 & 0 & 0 & 0 & 0 & 1\\
\multicolumn{1}{|c|}{OOOO} & 0 & 0 & 0 & 0 & 0 & 0.5 & 0 & 0.5 & 0 & 0 & 0 & 0.5\\
 \hline
    \end{tabular}
    \caption{Realized (contracted) matches in the case of Scenario III in the case of various strategy profiles (SPs).}
    \label{tab:Results_ex3_2_fullStratSpace}
\end{table*}

\begin{table*}[t]
    \centering
\begin{tabular}{|c|c|c|c|c|c|c|c|}
\hline
    \multicolumn{1}{|c|}{Strategy} & \multicolumn{1}{c|}{$U^C_1$\rule{0pt}{12pt}} & \multicolumn{1}{c|}{$U^C_2$} &
    \multicolumn{1}{c|}{$U^S_1$} & \multicolumn{1}{c|}{$U^S_2$} &
    \multicolumn{1}{c|}{TMQ} & \multicolumn{1}{c|}{TCQ} & \multicolumn{1}{c|}{TU} \\ \hline
NNNN & 2.4 & 2.4 & 0.6 & 0 & 2 & 2 & 5.4\\
ONNN & 4.4 & 2.4 & 0.45 & 0.3 & 3 & 2.5 & 7.55\\
NONN & 2.4 & 3.3 & 0.3 & 0.3 & 3 & 2 & 6.3\\
NNON & 2.4 & 2.4 & 0.6 & 0 & 2.5 & 2 & 5.4\\
NNNO & 4.025 & 2.4 & 0.6 & 0.125 & 2.5 & 2.5 & 7.15\\
OONN & 4 & 2.85 & 0.45 & 0.3 & 3 & 2.5 & 7.6\\
ONON & 4.4 & 2.4 & 0.45 & 0.3 & 4 & 2.5 & 7.55\\
ONNO & 1.6 & 2.4 & 0.3 & 0.15 & 4 & 1.5 & 4.45\\
NOON & 2.4 & 3.3 & 0.3 & 0.3 & 3.5 & 2 & 6.3\\
NONO & 2.4 & 3.3 & 0.3 & 0.3 & 3.5 & 2 & 6.3\\
NNOO & 4.025 & 2.4 & 0.6 & 0.125 & 2.5 & 2.5 & 7.15\\
OOON & 4 & 2.85 & 0.45 & 0.3 & 4 & 2.5 & 7.6\\
OONO & 1.6 & 2.85 & 0.15 & 0.3 & 4 & 1.5 & 4.9\\
ONOO & 1.6 & 2.4 & 0.3 & 0.15 & 4 & 1.5 & 4.45\\
NOOO & 2.4 & 3.3 & 0.3 & 0.3 & 3.5 & 2 & 6.3\\
OOOO & 1.6 & 2.85 & 0.15 & 0.3 & 4 & 1.5 & 4.9\\     \hline
\end{tabular}
    \caption{Utility values of individual players, total matched quantity (TMQ), total contracted quantity (TCQ) and total utility in the case of Scenario III for all strategy profiles}
    \label{tab:ex_3_utility_fullStratSpace}
\end{table*}



\end{document}